\documentclass[english]{article}
\usepackage[T1]{fontenc}
\usepackage{color}
\usepackage{graphicx}
\usepackage{amssymb}

\makeatletter

\usepackage{a4}
\input{epsfig.sty}
\def\fnum@table{\tablename~{\bf\thetable}}
\def\fnum@figure{\figurename~{\bf\thefigure}}
\def\tablename{\footnotesize{\bf Table}}
\def\figurename{\footnotesize{\bf Figure}}

\def\be{\begin{equation}}
\def\ee{\end{equation}}

\usepackage{babel}
\makeatother
\begin{document}

\title{\textbf{Non-linear screening effects }\\
\textbf{in high energy hadronic interactions} \textbf{\textsc{}}\\
}

\author{\textbf{S. Ostapchenko}%
 \textbf{}\\
\textit{\small Forschungszentrum Karlsruhe, Institut} \textit{\textcolor{black}{\small für}}
\textit{\small Kernphysik, 76021 Karlsruhe, Germany}\\
\textit{\small D.V. Skobeltsyn Institute of Nuclear Physics, Moscow
State University, 119992 Moscow, Russia} \textit{}\\
}

\maketitle
\begin{center}\textbf{\large Abstract}\end{center}{\large \par}

Non-linear effects in hadronic interactions are treated by means of
enhanced pomeron diagrams, assuming that pomeron-pomeron coupling
is dominated by soft partonic processes. It is shown that the approach
allows to resolve a seeming contradiction between realistic parton
momentum distributions, measured in deep inelastic scattering experiments,
and the energy behavior of total proton-proton cross section.  Also
a general consistency with both {}``soft'' and {}``hard'' diffraction
data is demonstrated. An important
feature of the proposed scheme is that the contribution of semi-hard
processes to the interaction eikonal contains a significant non-factorizable
part. On the other hand, the approach preserves the QCD factorization
picture for inclusive high-$p_{t}$ jet production.

\section{Introduction\label{intro.sec} }

One of the most important issues in high energy physics is the interplay
between soft and hard processes in hadronic interactions. The latter
involve parton evolution in the region of comparatively high virtualities
$q^{2}$ and can be treated within the perturbative QCD framework.
Despite the smallness of the running coupling $\alpha_{{\rm s}}(q^{2})$
involved, corresponding contributions are expected to dominate hadronic
interactions at sufficiently high energies, being enhanced by large
parton multiplicities and by large logarithmic ratios of parton transverse
and longitudinal momenta~\cite{glr}. On the other hand, very peripheral
hadronic collisions are likely to remain governed by non-perturbative
soft partonic processes, whose contribution to elastic scattering amplitude
thus remains significant even at very high energies. Furthermore,
considering production of high transverse momentum particles, one
may expect that a significant part of the underlying parton cascades,
which mediate the interaction, develops in the non-perturbative low
virtuality region \cite{dl94}, apart from the fact that additional
soft re-scattering processes may proceed in parallel to the mentioned
{}``semi-hard'' ones.

A popular scheme for a combined description of soft and hard processes
is the mini-jet approach \cite{minij}, employed in a
number of Monte Carlo generators \cite{minimc}. There,
one treats hadronic collisions within the eikonal framework, considering
the interaction eikonal to be the sum of {}``soft'' and {}``semi-hard''
contributions:\begin{eqnarray}
\chi_{ad}(s,b)=\frac{1}{2} \sigma_{ad}^{{\rm soft}}(s)\: A_{ad}(b)
+\frac{1}{2} \sigma_{ad}^{{\rm mini-jet}}(s,p_{t,{\rm min}})\: A_{ad}(b)\,,
\label{chi-mini-jet}\end{eqnarray}
where the overlap function $A_{ad}(b)$ is the convolution of electro-magnetic
form-factors of hadrons $a$ and $d$, $A_{ad}(b)=\int\! d^{2}b'\; T_{a}^{{\rm e/m}}(b')\; T_{d}^{{\rm e/m}}(|\vec{b}-\vec{b}'|)$,
$\sigma_{ad}^{{\rm soft}}(s)$ - a parameterized {}``soft'' parton
cross section, and $\sigma_{ad}^{{\rm mini-jet}}(s,p_{t,{\rm min}})$
is the inclusive cross section for production of parton jets with
transverse momentum bigger than a chosen cutoff $p_{t,{\rm min}}$,
for which the leading logarithmic QCD result is generally used.

Qualitatively similar is the {}``semi-hard pomeron'' scheme \cite{kal94,dre99},
where hadronic interactions are treated within the Gribov's reggeon
approach \cite{gri68,bak76} as multiple exchanges of
{}``soft'' and {}``semi-hard'' pomerons, the two objects corresponding
to soft and semi-hard re-scattering processes.

\begin{figure}[htb]
\begin{center}\includegraphics[%
  width=6cm,
  height=4cm]{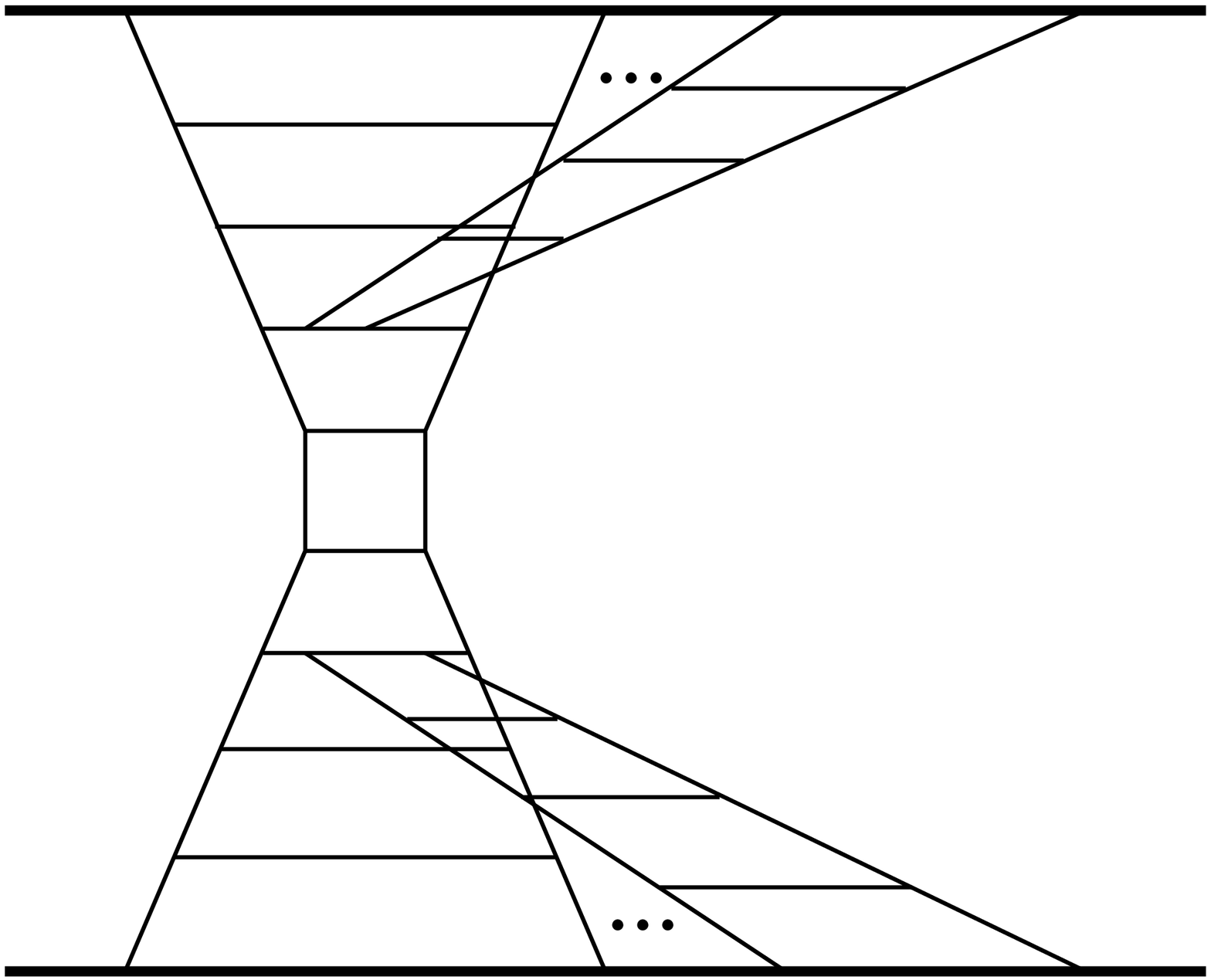}~~~~~~~~~~~~~~~\includegraphics[%
  width=6cm,
  height=4cm]{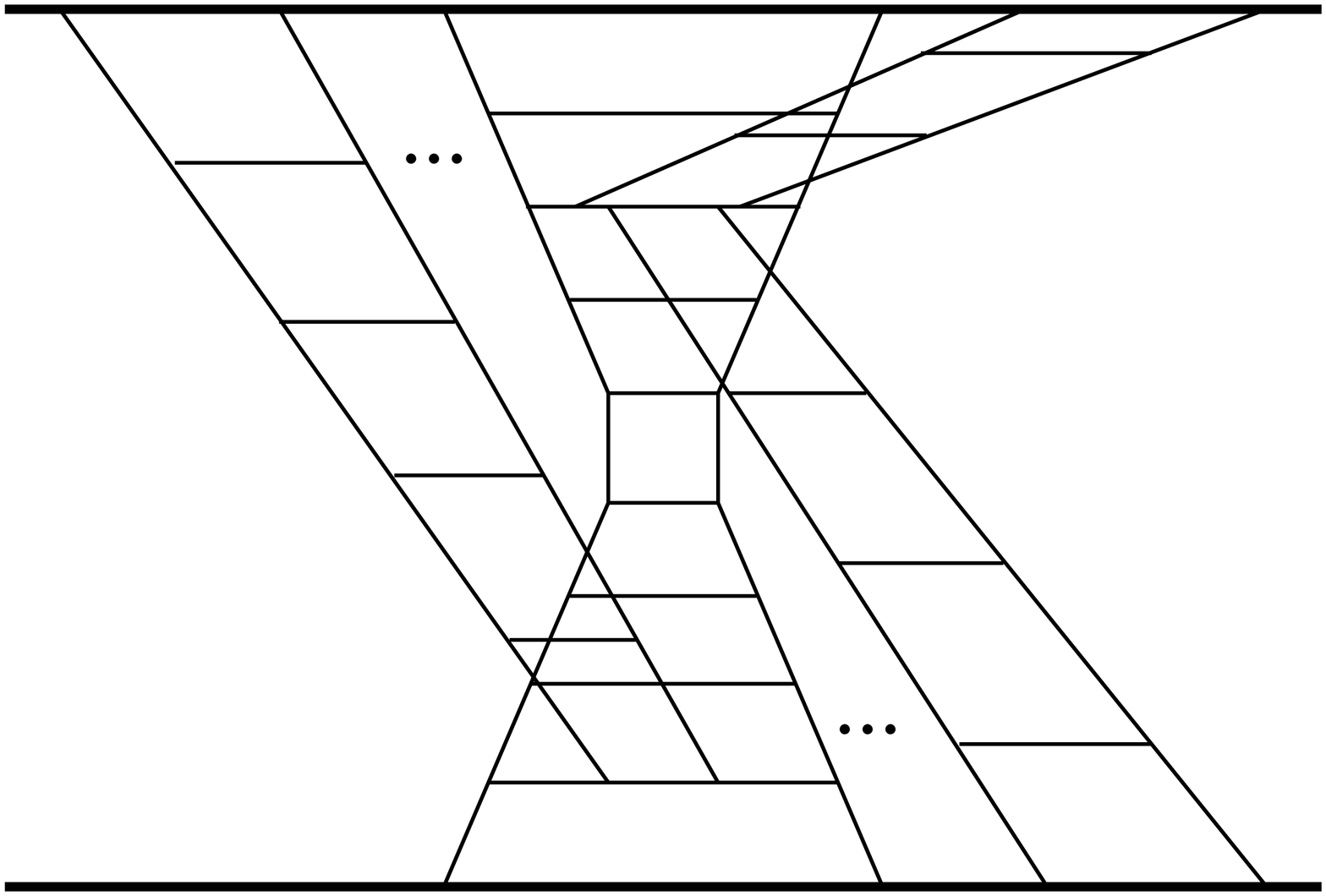}\end{center}
\vspace{-4mm}

\caption{Examples of diagrams giving rise to non-linear parton effects; increasing
parton virtualities are indicated as narrowing of the ladders.\label{ladder-merge}}
\end{figure}

However, both approaches face a fundamental difficulty when confronted
with available experimental data: it appears impossible to accommodate
realistic parton momentum distributions, when calculating the mini-jet
cross section $\sigma_{ad}^{{\rm mini-jet}}$ in (\ref{chi-mini-jet}),
without being in contradiction with moderately slow energy rise of
total proton-proton cross section. It seems natural to relate this
problem to the contribution of non-linear parton processes, which
are missing in the above-discussed eikonal scheme. Indeed, the need
for such corrections appears quite evident when considering small
$x$ behavior of parton momentum distribution functions (PDFs) at
some finite virtuality scale $Q^{2}$: due to a fast increase of,
e.g., gluon PDF $G(x,Q^{2})$ in the $x\rightarrow0$ limit parton
density in a restricted volume may reach arbitrarily large
 values \cite{glr}.
 In the QCD framework one describes non-linear parton
effects as merging of parton ladders, some typical
contributions shown in Fig.~\ref{ladder-merge}.%
In particular, one may account for such corrections in the perturbative
evolution of parton distributions, which leads to the saturation of
parton densities \cite{glr,mue86}.
 Moving towards smaller and smaller parton momentum fractions $x$, 
one obtains a higher saturation scale $Q_{{\rm sat}}^{2}(x)$,
with a dynamical parton evolution being only possible in the region
of sufficiently high virtualities $|q|^{2}>Q_{{\rm sat}}^{2}(x)$.

Using the mini-jet approach, one usually suggests some energy
dependence for the transverse momentum cutoff $p_{t,{\rm min}}$ for
mini-jet production, i.e.~$p_{t,{\rm min}}=p_{t,{\rm min}}(s)$, and
proceeds further with the usual eikonal expression (\ref{chi-mini-jet}).
There, the $p_{t}$-cutoff plays the role of an effective {}``saturation
scale'', for which a variety of empirical parameterizations has been
proposed \cite{boo94}. The underlying idea is to take
effectively into account the contributions of diagrams of Fig.~\ref{ladder-merge}(left),
where non-linear corrections (merging ladders) can be absorbed into
parton distributions. Unfortunately, introducing such empirical parameterizations
for the $p_{t}$-cutoff, one loses the connection to the perturbative
QCD and spoils predictive power of the method; the {}``saturation
scale'' is chosen irrespective the actual parton densities, which
depend on the parton momenta, on the {}``centrality'' of the interaction,
and on the projectile and target mass numbers in case of nuclear collisions.
On the other hand, there is no good reason for keeping the simple
relation (\ref{chi-mini-jet}) between the interaction eikonal and
the mini-jet cross section, when contributions of graphs of Fig.~\ref{ladder-merge}(right)
are taken into account.

In this paper a phenomenological treatment of non-linear screening
corrections is developed in the framework of Gribov's reggeon approach,
 describing the latter by means of enhanced (pomeron-pomeron
interaction) diagrams \cite{kan73,car74,bon01,ost06}. We employ
the {}``semi-hard pomeron'' approach,
taking into account both soft and semi-hard re-scattering processes.
 Assuming
that pomeron-pomeron coupling is dominated by non-perturbative soft
processes and using a phenomenological eikonal parameterization for 
multi-pomeron vertices,
 we  account for  enhanced corrections
to hadronic scattering amplitude and to total and diffractive structure
functions (SFs) $F_{2}$, $F_{2}^{D(3)}$. This allowed us to
obtain a consistent description of of total, elastic, and single diffraction
proton-proton cross sections, of the elastic scattering slope, and
of the SFs,
 using a fixed energy-independent virtuality cutoff
 for semi-hard processes. In particular,
we obtained significant corrections to the simple factorized expression
(\ref{chi-mini-jet}), which emerge from enhanced diagrams of the
kind of Fig.~\ref{ladder-merge}(right), where at least one pomeron
(additional parton ladder)
is exchanged in parallel to the hardest parton scattering process.
On the other hand, due to the Abramovskii-Gribov-Kancheli (AGK) cancellations
\cite{agk}, such diagrams do not contribute significantly to inclusive
parton spectra and the usual factorization picture remains applicable
for inclusive high $p_{t}$ jet production.

The outline of the paper is as follows. Section \ref{lin-scheme}
provides a brief overview of the {}``semi-hard pomeron'' approach.
Section \ref{sec:Non-linear-screening-corrections} is devoted to
the treatment of enhanced diagram contributions. Finally, the numerical
results obtained are discussed in Section \ref{sec:Results-and-discussion}.

\section{Linear scheme\label{lin-scheme}}

\begin{figure}[htb]
\begin{center}\includegraphics[%
  width=10cm,
  height=4cm]{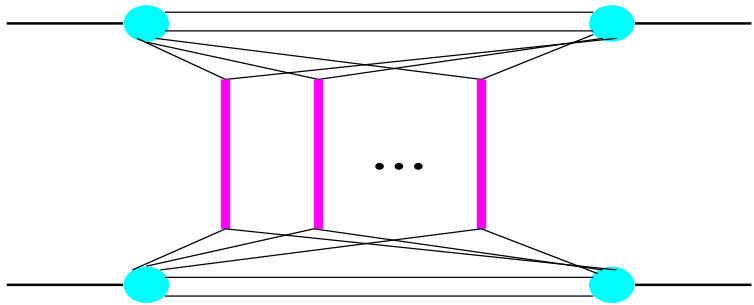}\end{center}
\vspace{-4mm}

\caption{A general multi-pomeron contribution to hadron-hadron scattering
amplitude; elementary scattering processes (vertical thick lines)
are described as pomeron exchanges.\label{multiple}}
\end{figure}

Using Gribov's reggeon approach \cite{gri68}, a high energy
hadron-hadron collision can be described as a multiple scattering
process, with elementary re-scatterings being treated phenomenologically
as pomeron exchanges, as shown in Fig.~\ref{multiple}.%
 Correspondingly, hadron $a$ - hadron $d$ elastic scattering amplitude
can be obtained summing over any number $n$ of pomeron exchanges%
\footnote{In the high energy limit all amplitudes can be considered as pure
imaginary.%
} \cite{bak76,bra90}:\begin{eqnarray}
i\, f_{ad}(s,b)=\sum_{j,k}C_{a(j)}\, C_{d(k)}\sum_{n=1}^{\infty}\frac{1}{n!}
\int\prod_{l=1}^{n}\left[dx_{l}^{+}dx_{l}^{-}\left(-\lambda_{a(j)}\,\lambda_{d(k)}
\, G_{ad}^{\mathbb{P}}(x_{l}^{+}x_{l}^{-}s,b)\right)\right]\nonumber \\
\times N_{a}^{(n)}(x_{1}^{+},\ldots,x_{n}^{+})\, N_{d}^{(n)}(x_{1}^{-},\ldots,x_{n}^{-})\,,
\label{f_ad}\end{eqnarray}
where $s$ and $b$ are c.m. energy squared and impact parameter for
the interaction, $G_{ad}^{\mathbb{P}}(x^{+}x^{-}s,b)$ is the un-integrated
pomeron exchange eikonal (for fixed values of pomeron light cone momentum
shares $x^{\pm}$), and $N_{a}^{(n)}(x_{1},\ldots,x_{n})$ is the
light cone momentum distribution of constituent partons - pomeron
{}``ends''. $C_{a(j)}$ and $\lambda_{a(j)}$ are correspondingly
relative weights and relative strengths of diffraction eigenstates
of hadron $a$ in Good-Walker formalism \cite{goo60}, $\sum_{j}C_{a(j)}=1$,
$\sum_{j}C_{a(j)}\lambda_{a(j)}=1$. In particular, two-component
picture ($j=1,2$) with one {}``passive'' component, $\lambda_{a(2)}\equiv0$,
corresponds to the usual quasi-eikonal approach \cite{kai82},
 with $\lambda_{a(1)}\equiv1/C_{a(1)}$
being the shower enhancement coefficient.

Assuming a factorized form%
\footnote{Here we neglect momentum correlations between multiple re-scattering
processes \cite{bra90}.%
} for $N_{a}^{(n)}$, i.e. $N_{a}^{(n)}(x_{1},\ldots,x_{n})=\prod_{i=1}^{n}N_{a}^{(1)}(x_{i})$,
one can simplify (\ref{f_ad}):\begin{eqnarray}
f_{ad}(s,b)=i\sum_{j,k}C_{a(j)}\, C_{d(k)}\left[1-e^{-\lambda_{a(j)}\,\lambda_{d(k)}\,\chi_{ad}^{\mathbb{P}}(s,b)}\right]\label{f_ad-fact}\\
\chi_{ad}^{\mathbb{P}}(s,b)=\int\! dx^{+}dx^{-}\, 
G_{ad}^{\mathbb{P}}(x^{+}x^{-}s,b)\, N_{a}^{(1)}(x^{+})\, N_{d}^{(1)}(x^{-})\,,
\label{chi_ad}\end{eqnarray}
where the vertex $N_{a}^{(1)}(x)$ can be parameterized as $N_{a}^{(1)}(x)\sim x^{-\alpha_{{\rm part}}}(1-x)^{\alpha_{a}^{{\rm lead}}}$,
with the parameters $\alpha_{{\rm part}}\simeq0$, $\alpha_{p}^{{\rm lead}}\simeq1.5$
related to intercepts of secondary reggeon trajectories \cite{kai82,dre01}.

This leads to traditional expressions for total and elastic cross
sections and for elastic scattering slope:
\begin{eqnarray}
\sigma_{ad}^{{\rm tot}}(s)=2\,{\rm Im}\int\!\! d^{2}b\, f_{ad}(s,b)=2\sum_{j,k}C_{a(j)}\, 
C_{d(k)}\int\!\! d^{2}b\;\left[1-e^{-\lambda_{a(j)}\,\lambda_{d(k)}\,
\chi_{ad}^{\mathbb{P}}(s,b)}\right]\label{sigma-tot}\\
\sigma_{ad}^{{\rm el}}(s)=\int\!\! d^{2}b\;\left[\sum_{j,k}C_{a(j)}\, C_{d(k)}\,
\left(1-e^{-\lambda_{a(j)}\,\lambda_{d(k)}\,\chi_{ad}^{\mathbb{P}}(s,b)}\right)\right]^{2}
\label{sigma-el}\\
B_{ad}^{{\rm el}}(s)=\left.\frac{d}{dt}\ln\frac{d\sigma_{ad}^{{\rm el}}(s,t)}{dt}\right|_{t=0}=
\frac{1}{\sigma_{ad}^{{\rm tot}}(s)}\sum_{j,k}C_{a(j)}\, C_{d(k)}\int\!\! d^{2}b\; b^{2}\;
\left[1-e^{-\lambda_{a(j)}\,\lambda_{d(k)}\,\chi_{ad}^{\mathbb{P}}(s,b)}\right],\label{b-el}
\end{eqnarray}
where $d\sigma_{ad}^{{\rm el}}(s,t)/dt$ is the differential elastic
cross section for momentum transfer squared $t$.

In turn, one obtains the cross section for low mass diffractive excitation of
the target hadron cutting the elastic scattering diagrams
of Fig.~\ref{multiple} in such a way that the cut plane passes between
uncut pomerons, with at least one remained on either side of the cut,
and selecting in the cut plane elastic intermediate state for hadron
$a$ and inelastic one for hadron $d$:\begin{eqnarray}
\sigma_{ad}^{{\rm LMD(targ)}}(s)=\int\!\! d^{2}b\;\sum_{j,k,l,m}\, C_{a(j)}\, C_{a(l)}\,\left(C_{d(k)}\,\delta_{k}^{m}-C_{d(k)}C_{d(m)}\right)\nonumber \\
\times\left(1-e^{-\lambda_{a(j)}\,\lambda_{d(k)}\,\chi_{ad}^{\mathbb{P}}(s,b)}\right)\;
\left(1-e^{-\lambda_{a(l)}\,\lambda_{d(m)}\,\chi_{ad}^{\mathbb{P}}(s,b)}\right).
\label{LMD-proj}\end{eqnarray}
The projectile single low mass diffraction cross section $\sigma_{ad}^{{\rm LMD(proj)}}(s)$
is obtained via the replacement $(a\longleftrightarrow d)$ in the
r.h.s.~of (\ref{LMD-proj}).

In this scheme the pomeron provides an effective description of a
microscopic parton cascade, which mediates the interaction between
the projectile and the target hadrons. At moderate energies the underlying
parton cascade for the pomeron exchange consists mainly of {}``soft''
partons of small virtualities and can be treated in a purely phenomenological
way. The corresponding eikonal can be chosen as \cite{bak76}\begin{eqnarray}
G_{ad}^{\mathbb{P}_{{\rm soft}}}(\hat{s},b)=\frac{\gamma_{0}^{2}\,\left(\hat{s}/s_{0}\right)^{\Delta}}{R_{a}^{2}+R_{d}^{2}+\alpha_{\mathbb{P}}'(0)\,\ln(\hat{s}/s_{0})}\;\exp\!\left[-\frac{b^{2}}{4\left(R_{a}^{2}+R_{d}^{2}+\alpha_{\mathbb{P}}'(0)\,\ln(\hat{s}/s_{0})\right)}\right],\label{chi-P}\end{eqnarray}
where $s_{0}\simeq1$ GeV$^{2}$ is the hadronic mass scale, $\Delta=\alpha_{\mathbb{P}}(0)-1$,
$\alpha_{\mathbb{P}}(0)$ and $\alpha_{\mathbb{P}}'(0)$ are the intercept
and the slope of the pomeron Regge trajectory, $R_{a}^{2}$ is the
Regge slope of hadron $a$, and $\gamma_{0}$ stands for pomeron coupling
to constituent partons. 

\begin{figure}[t]
\begin{center}\includegraphics[%
  width=10cm,
  height=4cm]{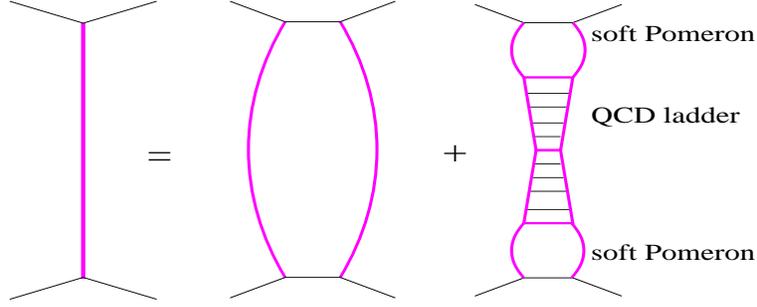}\end{center}
\vspace{-4mm}

\caption{A {}``general pomeron'' (l.h.s.) consists of the soft and semi-hard
ones - correspondingly the 1st and the 2nd contributions in the r.h.s.
\label{genpom} }
\end{figure}

At higher energies the underlying parton cascade is more and more
populated by quarks and gluons of comparatively high virtualities.
Dominant contribution comes here from hard scattering of gluons and
sea quarks, which are characterized by small fractions $x_{h}^{\pm}$
of parent hadron light cone momenta and are thus preceeded by extended
soft parton cascades ({}``soft pre-evolution''), covering long rapidity
intervals, $y_{{\rm soft}}\sim\ln1/x_{h}^{\pm}$ \cite{dl94}. One
may apply the phenomenological pomeron treatment for the low ($|q^{2}|<Q_{0}^{2}$
) virtuality part of the cascade and describe parton evolution at
higher virtualities $|q^{2}|>Q_{0}^{2}$ using pQCD techniques, $Q_{0}^{2}\sim1\div2$
GeV$^{2}$ being a reasonable scale for pQCD being applicable. Thus,
a cascade which at least partly develops in the high virtuality region
(some $|q^{2}|>Q_{0}^{2}$) can be described as an exchange of a {}``semi-hard
pomeron'', the latter being represented by a piece of QCD ladder
sandwiched between two soft pomerons%
\footnote{Similar approaches have been proposed in \cite{dl94,bon04}; 
in general, a {}``semi-hard pomeron'' may contain an arbitrary number
of $t$-channel iterations of soft and hard pomerons.
The word {}``pomeron'' appears here in quotes as the corresponding
amplitude is not the one of a Regge pole.%
} \cite{kal94,dre99}, see the 2nd graph in the r.h.s.~of Fig.~\ref{genpom}.
 Thus, the {}``general pomeron'' eikonal is the sum of soft
and semi-hard ones, as shown in Fig.~\ref{genpom}, and we
have \cite{dre99,dre01}\begin{eqnarray}
G_{ad}^{\mathbb{P}}(\hat{s},b)=G_{ad}^{\mathbb{P}_{{\rm soft}}}(\hat{s},b)+
G_{ad}^{\mathbb{P}_{{\rm sh}}}(\hat{s},b)\label{chi-tot}\\
G_{ad}^{\mathbb{P}_{{\rm sh}}}(\hat{s},b)=\frac{1}{2}\sum_{I,J=g,q_{s}}
\int\!\! d^{2}b'\int\!\frac{dx_{h}^{+}}{x_{h}^{+}}\frac{dx_{h}^{-}}{x_{h}^{-}}\; 
G_{aI}^{\mathbb{P}_{{\rm soft}}}\!\left(\frac{s_{0}}{x_{h}^{+}},b'\right)\, 
G_{dJ}^{\mathbb{P}_{{\rm soft}}}\!\left(\frac{s_{0}}{x_{h}^{-}},|\vec{b}-\vec{b}'|\right)\nonumber \\
\times\sigma_{IJ}^{{\rm QCD}}(x_{h}^{+}\, x_{h}^{-}\,\hat{s},Q_{0}^{2})\,.
\label{chi-sh}\end{eqnarray}
Here $\sigma_{IJ}^{{\rm QCD}}(x_{h}^{+}\, x_{h}^{-}\,\hat{s},Q_{0}^{2})$
stands for the contribution of parton ladder with the virtuality cutoff
$Q_{0}^{2}$; $I,J$ and $x_{h}^{+},x_{h}^{-}$ are types (gluons
and sea quarks) and relative light cone momentum fractions of ladder
leg partons:\begin{eqnarray}
\sigma_{IJ}^{{\rm QCD}}(\hat{s},Q_{0}^{2})=K\sum_{I',J'}\!\int\!\! dz^{+}dz^{-}\!\!
\int\!\! dp_{t}^{2}\; E_{I\rightarrow I'}^{{\rm QCD}}(z^{+},Q_{0}^{2},M_{{\rm F}}^{2})\, 
E_{J\rightarrow J'}^{{\rm QCD}}(z^{-},Q_{0}^{2},M_{{\rm F}}^{2})\nonumber \\
\times\frac{d\sigma_{I'J'}^{2\rightarrow2}(z^{+}z^{-}\hat{s},p_{t}^{2})}{dp_{t}^{2}}\:
\Theta(M_{F}^{2}-Q_{0}^{2})\,,\label{sigma-hard}\end{eqnarray}
where $d\sigma_{IJ}^{2\rightarrow2}/dp_{t}^{2}$ is the differential
parton-parton cross section, $p_{t}$ being the parton transverse
momentum in the hard process, $M_{F}^{2}$ - the factorization scale
(here $M_{F}^{2}=p_{t}^{2}/4$), the factor $K\simeq1.5$ takes effectively
into account higher order QCD corrections, and $E_{I\rightarrow I'}^{{\rm QCD}}(z,Q_{0}^{2},Q^{2})$
describes parton evolution from scale $Q_{0}^{2}$ to $Q^{2}$.

The eikonal $G_{aI}^{\mathbb{P}_{{\rm soft}}}(\hat{s},b)$, corresponding
to soft pomeron exchange between hadron $a$ and parton $I$, is obtained
from (\ref{chi-P}) replacing one vertex $\gamma_{0}$ by a parameterized
pomeron-parton vertex $\gamma_{I}(z),$ $z=s_{0}/\hat{s}$, and neglecting
a small slope of pomeron-parton coupling $R_{I}^{2}\sim1/Q_{0}^{2}$,
which gives\begin{eqnarray}
G_{aI}^{\mathbb{P}_{{\rm soft}}}(\hat{s},b)=\frac{\gamma_{0}\;\gamma_{I}(s_{0}/\hat{s})\;\left(\hat{s}/s_{0}\right)^{\Delta}}{R_{a}^{2}+\alpha_{\mathbb{P}}'(0)\,\ln(\hat{s}/s_{0})}\;\exp\!\left[-\frac{b^{2}}{4\left(R_{a}^{2}+\alpha_{\mathbb{P}}'(0)\,\ln(\hat{s}/s_{0})\right)}\right].\label{chi_aI}\end{eqnarray}
Here we use \cite{dre99,dre01}
\begin{eqnarray}
\gamma_{g}(z)=r_{g}\,(1-w_{qg})\;(1-z)^{\beta_{g}}\label{gamma-g}\\
\gamma_{q_{s}}(z)=r_{g}\,w_{qg}\int_{z}^{1}\! dy\; y^{\Delta}\; 
P_{qg}(y)\;(1-z/y)^{\beta_{g}},\label{gamma-q}
\end{eqnarray}
where $P_{qg}(y)=3\,[y^{2}+(1-y)^{2}]$ is the usual Altarelli-Parisi
splitting kernel for three active flavors. By construction, the eikonal
$G_{aI}^{\mathbb{P}_{{\rm soft}}}(s_{0}/x,b)$ describes momentum
fraction $x$ and impact parameter $b$ distribution of parton $I$
(gluon or sea quark) in the soft pomeron at virtuality scale $Q_{0}^{2}$,
with the constant $r_{g}$ being fixed by parton momentum conservation
\begin{eqnarray}
\int_{0}^{1}\! dx\int\! d^{2}b\;\left[G_{ag}^{\mathbb{P}_{{\rm soft}}}(s_{0}/x,b)+
G_{aq_{s}}^{\mathbb{P}_{{\rm soft}}}(s_{0}/x,b)\right]=1\,.\label{mom-cons}\end{eqnarray}

Convoluting $G_{aI}^{\mathbb{P}_{{\rm soft}}}$ with the constituent
parton distribution $N_{a}^{(1)}(x)$, one obtains momentum and impact
parameter distribution of parton $I$ in hadron $a$ at virtuality
scale $Q_{0}^{2}$: \begin{eqnarray}
x\,\tilde{f}_{I/a}(x,b,Q_{0}^{2})=\int_{x}^{1}\! dx'\, N_{a}^{(1)}(x')\; G_{aI}^{\mathbb{P}_{{\rm soft}}}\!\left(\frac{s_{0}\, x'}{x},b\right).\label{sf-q0}\end{eqnarray}

In addition to $\chi_{ad}^{\mathbb{P}}(s,b)$, defined by (\ref{chi_ad}),
(\ref{chi-P}--\ref{chi-sh}), one may include contributions of valence
quark hard interactions with each other or with sea quarks and gluons%
\footnote{For brevity, in the following these contributions will not be discussed
explicitely. As will be shown below, the predictions for high energy
hadronic cross sections depend rather weakly on the input valence
quark PDFs.%
} $\chi_{ad}^{{\rm val-val}}$, $\chi_{ad}^{{\rm val-sea}}$, $\chi_{ad}^{{\rm sea-val}}$
\cite{dre99,dre01}. In case of valence quarks one can neglect
the {}``soft pre-evolution'' and use for their momentum and impact
parameter distribution at scale $Q_{0}^{2}$\begin{eqnarray}
\tilde{f}_{q_{v}/a}(x,b,Q_{0}^{2})=\frac{q_{v}(x,Q_{0}^{2})}{4\pi\, R_{a}^{2}}\;\exp\!\left(-\frac{b^{2}}{4R_{a}^{2}}\right),\label{sf-val}\end{eqnarray}
with $q_{v}(x,Q_{0}^{2})$ being a parameterized input (here GRV94
\cite{grv94}).

Correspondingly, the complete hadron-hadron interaction eikonal can
be written as \cite{dre99,dre01}\begin{eqnarray}
\chi_{ad}(s,b)=\chi_{ad}^{\mathbb{P}}(s,b)+\chi_{ad}^{{\rm val-val}}(s,b)+
\chi_{ad}^{{\rm val-sea}}(s,b)+\chi_{ad}^{{\rm sea-val}}(s,b)\nonumber \\
=\chi_{ad}^{\mathbb{P}_{{\rm soft}}}(s,b)+\frac{K}{2}\sum_{I,J}\!\int\!\! dx^{+}dx^{-}\!
\int\!\! dp_{t}^{2}\!\int\!\! d^{2}b'\;\tilde{f}_{I/a}(x^{+},b',M_{F}^{2})\:
\tilde{f}_{J/d}(x^{-},|\vec{b}-\vec{b}'|,M_{F}^{2})\nonumber \\
\times\frac{d\sigma_{IJ}^{2\rightarrow2}(x^{+}x^{-}s,p_{t}^{2})}{dp_{t}^{2}}\:
\Theta(M_{F}^{2}-Q_{0}^{2})\,,\label{chi-total}\end{eqnarray}
where $\chi_{ad}^{\mathbb{P}_{{\rm soft}}}(s,b)=\int\! dx^{+}dx^{-}\, G_{ad}^{\mathbb{P}_{{\rm soft}}}(x^{+}x^{-}s,b)\, N_{a}^{(1)}(x^{+})\, N_{d}^{(1)}(x^{-})$
and parton momentum and impact parameter distributions $\tilde{f}_{I/a}(x,b,Q^{2})$
at arbitrary scale $Q^{2}$ are obtained evolving the input ones (\ref{sf-q0}-\ref{sf-val})
from $Q_{0}^{2}$ to $Q^{2}$: 
\begin{eqnarray}
\tilde{f}_{I/a}(x,b,Q^{2})=\sum_{J=g,q_{s},q_{v}}\!\int_{x}^{1}
\!\!\frac{dz}{z}\: E_{J\rightarrow I}^{{\rm QCD}}(z,Q_{0}^{2},Q^{2})\;
\tilde{f}_{J/a}(x/z,b,Q_{0}^{2})\,.\label{pdf}\end{eqnarray}

It is noteworthy that the eikonal (\ref{chi-total}) is similar to
the usual ansatz (\ref{chi-mini-jet}) of the mini-jet approach, apart
from the fact that in the latter case one assumed a factorized momentum
and impact parameter dependence of parton distributions, i.e.\begin{eqnarray}
\tilde{f}_{I/a}^{{\rm mini-jet}}(x,b,Q^{2})=f_{I/a}(x,Q^{2})\; T_{a}^{{\rm e/m}}(b)\,.
\label{pdf-fact}\end{eqnarray}
In the above-described approach parton distributions at arbitrary
scale $Q^{2}$ are obtained from a convolution of {}``soft'' and
{}``hard'' parton evolution, the former being described by the soft
pomeron asymptotics. As a consequence, partons of smaller virtualities
result from a longer {}``soft'' evolution and are distributed over
a larger transverse area. On the other hand, the latter circumstance
is closely related to the chosen functional form (\ref{chi-P}) for
the pomeron amplitude, characterized by a Gaussian impact parameter
dependence. In the mini-jet approach one typically employs a dipole
parameterization for hadronic form-factors $T_{a}^{{\rm e/m}}(b)$,
which allows to put the slope of the soft contribution down to zero
and thus leads to the geometrical scaling picture.

\section{Non-linear screening corrections\label{sec:Non-linear-screening-corrections}}

The above-described picture appears to be incomplete in the {}``dense''
regime, i.e.~in the limit of high energies and small impact parameters
for the interaction. There, a large number of elementary scattering
processes occurs and corresponding underlying parton cascades overlap
and interact with each other, giving rise to significant non-linear
effects. Here we are going to treat non-linear screening
effects in the framework of Gribov's reggeon scheme \cite{gri68,bak76}
by means of enhanced pomeron diagrams, which involve pomeron-pomeron
interactions \cite{kan73,car74}. Concerning multi-pomeron
vertices, we assume  that they are characterized by small slope
$R_{\mathbb{P}}^{2}$ (neglected in the following) and by eikonal
structure, i.e.~for the vertex which describes the transition of
$m$ into $n$ pomerons we use 
\begin{eqnarray}
g_{mn}=r_{3\mathbb{P}}\,\gamma_{\mathbb{P}}^{m+n-3}/(4\pi\, m!\, n!)\,,
\label{g_mn}\end{eqnarray}
with $r_{3\mathbb{P}}$ being the triple-pomeron coupling. Doing a
replacement $r_{3\mathbb{P}}=4\pi\, G\,\gamma_{\mathbb{P}}^{3}$ and
neglecting momentum spread of pomeron {}``ends'' in the vertices,
for a pomeron exchanged between two vertices, separated from each
other by rapidity $y$ and impact parameter $b$, we use the eikonal
$G_{\mathbb{PP}}^{\mathbb{P}}(y,b)$, being the sum of corresponding
soft and semi-hard contributions $G_{\mathbb{PP}}^{\mathbb{P}_{{\rm soft}}}(y,b)$,
$G_{\mathbb{PP}}^{\mathbb{P}_{{\rm sh}}}(y,b)$. The latter are obtained
from $G_{ad}^{\mathbb{P}_{{\rm soft}}}(s_{0}\, e^{y},b)$, $G_{ad}^{\mathbb{P}_{{\rm sh}}}(s_{0}\, e^{y},b)$,
defined in (\ref{chi-P}), (\ref{chi-sh}-\ref{chi_aI}), replacing
the vertex factors $\gamma_{a}$, $\gamma_{d}$ by $\gamma_{\mathbb{P}}$
and the slopes $R_{a}^{2}$, $R_{d}^{2}$ by $R_{\mathbb{P}}^{2}\sim0$:\begin{eqnarray}
G_{\mathbb{PP}}^{\mathbb{P}}(y,b)=G_{\mathbb{PP}}^{\mathbb{P}_{{\rm soft}}}(y,b)+G_{\mathbb{PP}}^{\mathbb{P}_{{\rm sh}}}(y,b)\label{chi-PPP}\\
G_{\mathbb{PP}}^{\mathbb{P}_{{\rm soft}}}(y,b)=\frac{\gamma_{\mathbb{P}}^{2}\; e^{\Delta\, y}}{\alpha_{\mathbb{P}}'(0)\, y}\;\exp\!\left[-\frac{b^{2}}{4\alpha_{\mathbb{P}}'(0)\, y}\right]\label{PPP-soft}\\
G_{\mathbb{PP}}^{\mathbb{P}_{{\rm sh}}}(y,b)=\frac{1}{2}\sum_{I,J}\int\!\! d^{2}b'\!\int\! dy^{+}\, dy^{-}\; G_{\mathbb{P}I}^{\mathbb{P}_{{\rm soft}}}(y^{+},b')\, G_{\mathbb{P}J}^{\mathbb{P}_{{\rm soft}}}(y^{-},|\vec{b}-\vec{b}'|)\,\sigma_{IJ}^{{\rm QCD}}(s_{0}\, e^{y-y^{+}-y^{-}},Q_{0}^{2})\label{PPP-sh}\\
G_{\mathbb{P}I}^{\mathbb{P}_{{\rm soft}}}(y,b)=\frac{\gamma_{\mathbb{P}}\;\gamma_{I}(e^{-y})\; e^{\Delta\, y}}{\alpha_{\mathbb{P}}'(0)\, y}\,\exp\!\left[-\frac{b^{2}}{4\alpha_{\mathbb{P}}'(0)\, y}\right].\label{PPI}\end{eqnarray}

Similarly, for a pomeron exchanged between hadron $a$ and a multi-pomeron
vertex we use the eikonal $\chi_{a\mathbb{P}}^{\mathbb{P}}(y,b)$,
defined as\begin{eqnarray}
\chi_{a\mathbb{P}}^{\mathbb{P}}(y,b)=\int\! dx\; N_{a}^{(1)}(x)\,\left[G_{a\mathbb{P}}^{\mathbb{P}_{{\rm soft}}}(y-\ln\!\frac{1}{x},b)+G_{a\mathbb{P}}^{\mathbb{P}_{{\rm sh}}}(y-\ln\!\frac{1}{x},b)\right]\label{chi-aPP}\\
G_{a\mathbb{P}}^{\mathbb{P}_{{\rm soft}}}(y,b)=\frac{\gamma_{0}\;\gamma_{\mathbb{P}}\; e^{\Delta\, y}}{R_{a}^{2}+\alpha_{\mathbb{P}}'(0)\, y}\;\exp\!\left[-\frac{b^{2}}{4\left(R_{a}^{2}+\alpha_{\mathbb{P}}'(0)\, y\right)}\right]\label{aPP-soft}\\
G_{a\mathbb{P}}^{\mathbb{P}_{{\rm sh}}}(y,b)=
\frac{1}{2}\sum_{I,J}\int\!\! d^{2}b'\!\int\! dy^{+}\, dy^{-}\; 
G_{aI}^{\mathbb{P}_{{\rm soft}}}(y^{+},b')\; 
G_{\mathbb{P}J}^{\mathbb{P}_{{\rm soft}}}(y^{-},|\vec{b}-\vec{b}'|)\;
\sigma_{IJ}^{{\rm QCD}}\!(s_{0}\, e^{y-y^{+}-y^{-}},Q_{0}^{2})\,.\label{aPP-sh}\end{eqnarray}

\begin{figure}[t]
\begin{center}\includegraphics[%
  width=6cm,
  height=4cm]{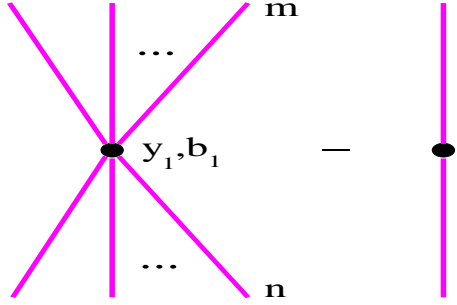}\end{center}
\vspace{-4mm}

\caption{Lowest order enhanced graphs; pomeron connections to the projectile
and target hadrons not shown explicitely.\label{PPP-1}}
\end{figure}

As an example, the contribution of enhanced diagrams with only one
multi-pomeron vertex, which are coupled to diffractive eigenstates
$j$ and $k$ of hadrons $a$ and $d$, can be obtained using standard
reggeon calculus techniques \cite{gri68,bak76,kan73,car74}: summing over
$m\geq1$ pomerons exchanged between the vertex and the projectile
hadron, $n\geq1$ pomeron exchanges between the vertex and the target,
subtracting the term with $m=n=1$ (pomeron self-coupling), and integrating
over rapidity $y_{1}<Y=\ln\frac{s}{s_{0}}$ and impact parameter $\vec{b}_{1}$
of the vertex, as shown in Fig.~\ref{PPP-1}:\begin{eqnarray}
\chi_{ad(jk)}^{\mathbb{PPP}(1)}\!(s,b)=\frac{G}{\lambda_{a(j)}\,\lambda_{d(k)}}
\sum_{m,n\geq1;m+n\geq3}\int_{0}^{Y}\!\! dy_{1}\!\int\!\! d^{2}b_{1}\;
\frac{\left[-\lambda_{a(j)}\,
\chi_{a\mathbb{P}}^{\mathbb{P}}(Y-y_{1},|\vec{b}-\vec{b}_{1}|)\right]^{m}}{m!}
\nonumber \\
\times\frac{\left[-\lambda_{d(k)}\,\chi_{d\mathbb{P}}^{\mathbb{P}}(y_{1},b_{1})
\right]^{n}}{n!}=\frac{G}{\lambda_{a(j)}\,\lambda_{d(k)}}
\int_{0}^{Y}\!\! dy_{1}\!\int\!\! d^{2}b_{1}\,\left\{ \left(1-e^{-\lambda_{a(j)}\,\chi_{a\mathbb{P}}^{\mathbb{P}}(Y-y_{1},|\vec{b}-\vec{b}_{1}|)}\right)\right.\nonumber \\
\left.\times\left(1-e^{-\lambda_{d(k)}\,\chi_{d\mathbb{P}}^{\mathbb{P}}(y_{1},b_{1})}\right)-\lambda_{a(j)}\,\lambda_{d(k)}\,\chi_{a\mathbb{P}}^{\mathbb{P}}(Y-y_{1},|\vec{b}-\vec{b}_{1}|)\;\chi_{d\mathbb{P}}^{\mathbb{P}}(y_{1},b_{1})\right\} .\label{3P-1}\end{eqnarray}

\begin{figure}[htb]
\begin{center}\includegraphics[%
  width=12cm,
  height=4cm]{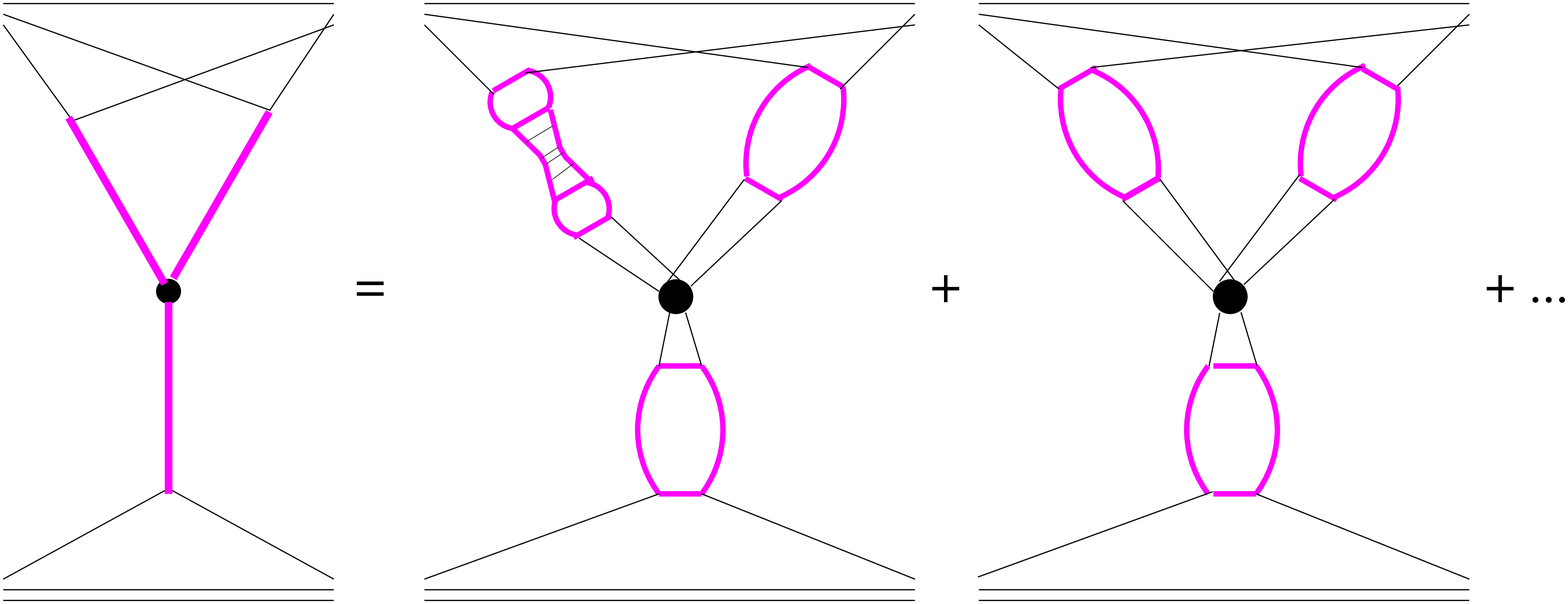}\end{center}
\vspace{-4mm}

\caption{Contributions to the triple-pomeron vertex from interactions between
soft and semi-hard pomerons.\label{3p-vertex}}
\end{figure}

Here our key assumption is that pomeron-pomeron coupling proceeds
via partonic processes at comparatively low virtualities, $|q^{2}|<Q_{0}^{2}$,
with $Q_{0}$ being a fixed energy-independent parameter \cite{dre01,ost05}.
In that case multi-pomeron vertexes involve only interactions between
soft pomerons or between  {}``soft ends'' of semi-hard pomerons,
as shown in Fig.~\ref{3p-vertex}; direct coupling between parton
ladders in the region of high virtualities $|q^{2}|>Q_{0}^{2}$ is
neglected. %

\begin{figure}[htb]
\begin{center}\includegraphics[%
  width=15cm,
  height=4cm]{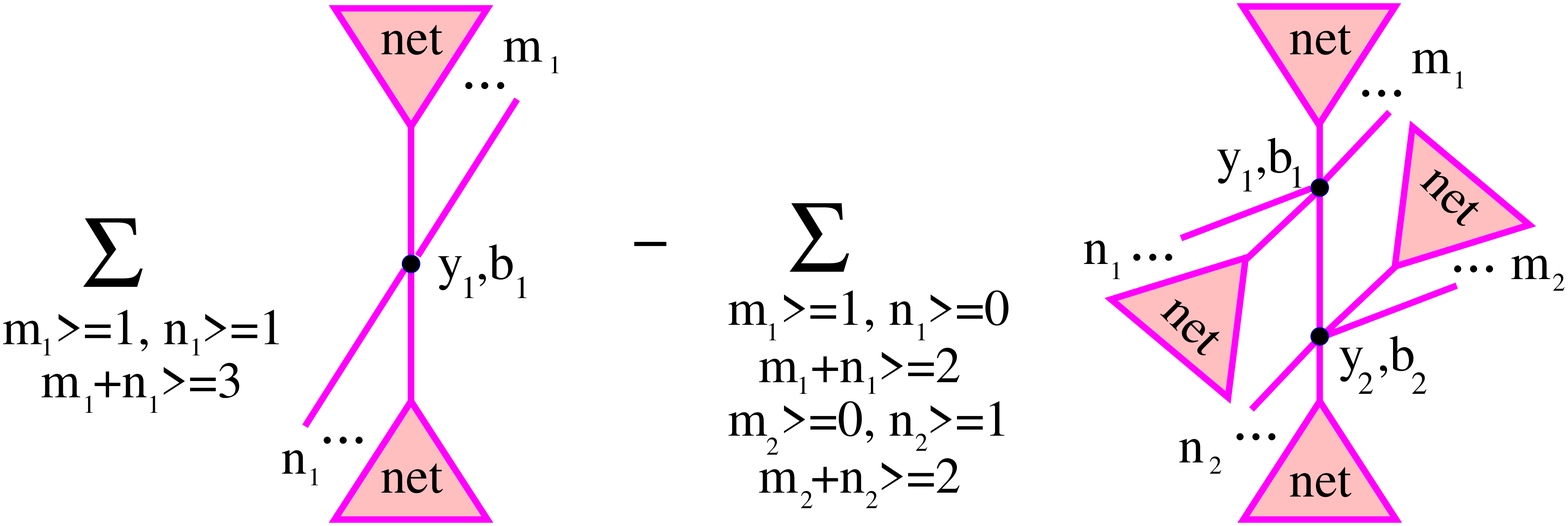}\end{center}
\vspace{-4mm}

\caption{Complete set of dominant enhanced diagrams; $y_{i}$, $\vec{b}_{i}$
($i=1,2$) denote rapidity and impact parameter positions of multi-pomeron
vertices, $i$-th vertex couples together $m_{i}$ projectile and
$n_{i}$ target {}``net fans''.\label{enh-full}}
\end{figure}

As shown in \cite{ost06}, the contribution of dominant enhanced diagrams
can be represented by the graphs of Fig.~\ref{enh-full}.%
 With our present conventions the corresponding eikonal contribution
can be written as%
\footnote{The expression for $\chi_{ad}^{{\rm enh}}(s,b)$ in \cite{ost06}
corresponds to the quasi-eikonal approach and to the $\pi$-meson
dominance of multi-pomeron vertices. %
} \begin{eqnarray}
\chi_{ad(jk)}^{{\rm enh}}(s,b)=\frac{G}{\lambda_{a(j)}\,\lambda_{d(k)}}\int_{0}^{Y}\!\! dy_{1}\!\int\!\! d^{2}b_{1}\left\{ \left[\left(1-e^{-\lambda_{a(j)}\,\chi_{a(j)|d(k)}^{{\rm net}}\!(Y-y_{1},\vec{b}-\vec{b}_{1}|Y,\vec{b})}\right)\right.\right.\nonumber \\
\left.\times\left(1-e^{-\lambda_{d(k)}\,\chi_{d(k)|a(j)}^{{\rm net}}\!(y_{1},\vec{b}_{1}|Y,\vec{b}}\right)-\lambda_{a(j)}\lambda_{d(k)}\,\chi_{a(j)|d(k)}^{{\rm net}}\!(Y-y_{1},\vec{b}-\vec{b}_{1}|Y,\vec{b})\;\chi_{d(k)|a(j)}^{{\rm net}}\!(y_{1},\vec{b}_{1}|Y,\vec{b})\right]\nonumber \\
-G\int_{0}^{y_{1}}\!\! dy_{2}\!\int\!\! d^{2}b_{2}\; G_{\mathbb{PP}}^{\mathbb{P}}(y_{1}-y_{2},|\vec{b}_{1}-\vec{b}_{2}|)\;\left[\left(1-e^{-\lambda_{a(j)}\,\chi_{a(j)|d(k)}^{{\rm net}}\!(Y-y_{1},\vec{b}-\vec{b}_{1}|Y,\vec{b})}\right)\right.\nonumber \\
\left.\times e^{-\lambda_{d(k)}\,\chi_{d(k)|a(j)}^{{\rm net}}\!(y_{1},\vec{b}_{1}|Y,\vec{b})}-\lambda_{a(j)}\,\chi_{a(j)|d(k)}^{{\rm net}}\!(Y-y_{1},\vec{b}-\vec{b}_{1}|Y,\vec{b})\right]\nonumber \\
\left.\times\left[\left(1-e^{-\lambda_{d(k)}\,\chi_{d(k)|a(j)}^{{\rm net}}\!(y_{2},\vec{b}_{2}|Y,\vec{b})}\right)e^{-\lambda_{a(j)}\,\chi_{a(j)|d(k)}^{{\rm net}}\!(Y-y_{2},\vec{b}-\vec{b}_{2}|Y,\vec{b})}-\lambda_{d(k)}\,\chi_{d(k)|a(j)}^{{\rm net}}\!(y_{2},\vec{b}_{2}|Y,\vec{b})\right]\right\} .\label{enh-tot}\end{eqnarray}
Here $\chi_{a(j)|d(k)}^{{\rm net}}\!(y,\vec{b}_{1}|Y,\vec{b})$ stands
for the contribution of {}``net fan'' graphs, which correspond to
arbitrary {}``nets'' of pomerons, exchanged between hadrons $a$
and $d$ (represented by their diffractive components $j,k$), with
one pomeron vertex in the {}``handle'' of the {}``fan'' being
fixed; $y$, $b_{1}$ are rapidity and impact parameter distances
between hadron $a$ and this vertex. The {}``net fan'' contribution
$\chi_{a(j)|d(k)}^{{\rm net}}$ is defined via the recursive equation
of Fig.~\ref{freve} \cite{ost06}:%
\begin{figure}[t]
\begin{center}\includegraphics[%
  width=12cm,
  height=4cm]{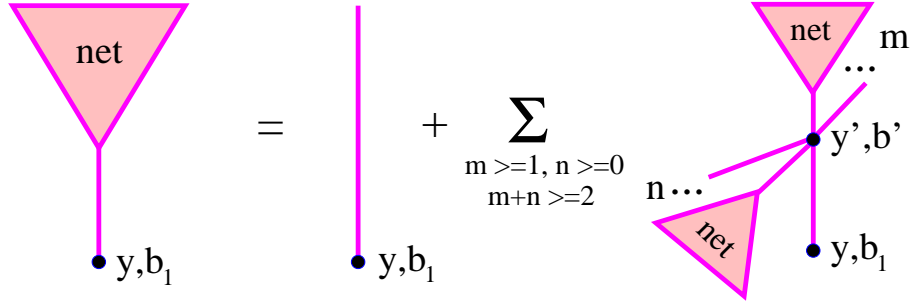}\end{center}
\vspace{-4mm}

\caption{Recursive equation for the projectile {}``net fan'' contribution
$\chi_{a(j)|d(k)}^{{\rm net}}\!(y,\vec{b}_{1}|Y,\vec{b})$; $y$,
$b_{1}$ are rapidity and impact parameter distances between hadron
$a$ and the vertex in the {}``handle'' of the {}``fan''. The
vertex $(y',b')$ couples together $m$ projectile and $n$ target
{}``net fans''.\label{freve}}
\end{figure}
 \begin{eqnarray}
\chi_{a(j)|d(k)}^{{\rm net}}(y,\vec{b}_{1}|Y,\vec{b})=\chi_{a\mathbb{P}}^{\mathbb{P}}(y,b_{1})+\frac{G}{\lambda_{a(j)}}\int_{0}^{y}\!\! dy'\int\!\! d^{2}b'\; G_{\mathbb{PP}}^{\mathbb{P}}(y-y',|\vec{b}_{1}-\vec{b}'|)\nonumber \\
\times\left[\left(1-e^{-\lambda_{a(j)}\,\chi_{a(j)|d(k)}^{{\rm net}}\!(y',\vec{b}'|Y,\vec{b})}\right)\, e^{-\lambda_{d(k)}\,\chi_{d(k)|a(j)}^{{\rm net}}\!(Y-y',\vec{b}-\vec{b}'|Y,\vec{b})}-\lambda_{a(j)}\,\chi_{a(j)|d(k)}^{{\rm net}}\!(y',\vec{b}'|Y,\vec{b})\right].\label{net-fan}\end{eqnarray}

Thus, one can calculate total, elastic, and single low mass diffraction
cross sections, as well as the elastic scattering slope for hadron-hadron
scattering, with non-linear screening corrections taken into account,
using usual expressions (\ref{sigma-tot}-\ref{LMD-proj}), with the
pomeron eikonal $\chi_{ad}^{\mathbb{P}}$ being replaced by the sum
of $\chi_{ad}^{\mathbb{P}}$ and $\chi_{ad(jk)}^{{\rm enh}}$:\begin{eqnarray}
\chi_{ad(jk)}^{{\rm tot}}(s,b)=\chi_{ad}^{\mathbb{P}}(s,b)+
\chi_{ad(jk)}^{{\rm enh}}(s,b)\,.\label{chi_ad-tot}\end{eqnarray}
In addition, considering different unitarity cuts of elastic scattering
diagrams of Figs.~\ref{multiple},~\ref{enh-full}, one can obtain
cross sections for various inelastic final states in hadron-hadron
interactions, including ones characterized by a rapidity gap signature.
While a general analysis of that kind is beyond the scope of the current
work and will be presented elsewhere \cite{ost07}, we include in
the Appendix a simplified derivation of single high mass diffraction
 cross section, with the final result being defined by
  (\ref{2-gap}-\ref{sig-hmd}).

\begin{figure}[t]
\begin{center}\includegraphics[%
  width=10cm,
  height=4cm]{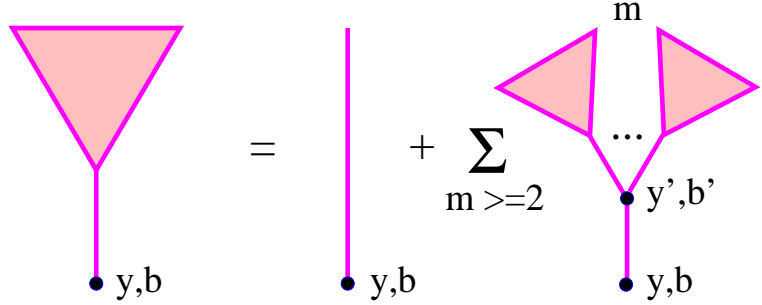}\end{center}
\vspace{-4mm}

\caption{Recursive equation for the projectile {}``fan'' contribution $\chi_{a(j)}^{{\rm fan}}(y,b)$;
$y$ and $b$ are rapidity and impact parameter distances between
hadron $a$ and the vertex in the {}``handle'' of the {}``fan''.
The vertex $(y',b')$ couples together $m$ projectile {}``fans''.\label{ffan}}
\end{figure}

Let us also derive screening corrections to parton (sea quark and
gluon) momentum and impact parameter distributions $\tilde{f}_{I/a}(x,b,Q^{2})$,
which come from diagrams of {}``fan'' type \cite{glr}. In our scheme
the general {}``fan'' contribution can be obtained solving iteratively
the recursive equation of Fig.~\ref{ffan},%
 which is a particular case of more general {}``net fan'' equation
of Fig.~\ref{freve}, when all intermediate vertices are connected
to hadron $a$ only (i.e.~$n\equiv0$ in Fig.~\ref{freve}):\begin{eqnarray}
\chi_{a(j)}^{{\rm fan}}(y,b)=\chi_{a\mathbb{P}}^{\mathbb{P}}(y,b)+\frac{G}{\lambda_{a(j)}}\int_{0}^{y}\!\! dy'\int\!\! d^{2}b'\; G_{\mathbb{PP}}^{\mathbb{P}}(y-y',|\vec{b}-\vec{b}'|)\nonumber \\
\times\left[1-e^{-\lambda_{a(j)}\,\chi_{a(j)}^{{\rm fan}}(y',b')}-\lambda_{a(j)}\,\chi_{a(j)}^{{\rm fan}}(y',b')\right].\label{fan}\end{eqnarray}

Then, parton distributions $x\,\tilde{f}_{I/a}^{{\rm scr}}(x,b,Q_{0}^{2})$
($I=g,q_s$) are defined by diagrams of Fig.~\ref{ffan} with $y=-\ln x$ and
with the down-most vertices being replaced by the pomeron-parton coupling,
which amounts to replace the eikonals $\chi_{a\mathbb{P}}^{\mathbb{P}}(y,b)$,
$G_{\mathbb{PP}}^{\mathbb{P}}(y-y',|\vec{b}-\vec{b}'|)$ in (\ref{fan})
by $x\,\tilde{f}_{I/a}(x,b,Q_{0}^{2})$, $G_{\mathbb{P}I}^{\mathbb{P}_{{\rm soft}}}(-\ln x-y',|\vec{b}-\vec{b}'|)$
correspondingly, the latter being defined in (\ref{sf-q0}), (\ref{PPI}).
Thus, averaging over diffraction eigenstates of hadron $a$ with the
corresponding weights $C_{a(j)}\lambda_{a(j)}$, we obtain\begin{eqnarray}
x\,\tilde{f}_{I/a}^{{\rm scr}}(x,b,Q_{0}^{2})=x\,\tilde{f}_{I/a}(x,b,Q_{0}^{2})+G\sum_{j}C_{a(j)}\int_{0}^{-\ln x}\!\! dy'\int\!\! d^{2}b'\; G_{\mathbb{P}I}^{\mathbb{P}_{{\rm soft}}}(-\ln x-y',|\vec{b}-\vec{b}'|)\nonumber \\
\times\left[1-e^{-\lambda_{a(j)}\,\chi_{a(j)}^{{\rm fan}}(y',b')}-\lambda_{a(j)}\,\chi_{a(j)}^{{\rm fan}}(y',b')\right].
\label{pdf-scr}\end{eqnarray}
Parton distributions $\tilde{f}_{I/a}^{{\rm scr}}(x,b,Q^{2})$ at
arbitrary scale $Q^{2}$ are obtained substituting $\tilde{f}_{J/a}(x,b,Q_{0}^{2})$
in (\ref{pdf}) by $\tilde{f}_{J/a}^{{\rm scr}}(x,b,Q_{0}^{2})$ as
the initial conditions for sea quarks and gluons.

Let us finally obtain diffractive parton distributions, which are
relevant for diffractive deep inelastic scattering reactions, when
a large rapidity gap, not covered by secondary particle production,
appears in the process. First, we have to obtain the contribution
of unitarity cuts of the {}``fan'' diagrams of Fig.~\ref{ffan},
which lead to a rapidity gap of size $y_{{\rm gap}}$ between hadron
$a$ and the nearest particle produced after the gap. Introducing
a generic symbol for the  diffractive contribution 
 $2\chi_{a(j)}^{{\rm diffr}}(y,b,y_{{\rm gap}})$ 
as a {}``fork'' with broken {}``handle'', applying AGK cutting
rules \cite{agk} to the 2nd graph in the r.h.s.~of Fig.~\ref{ffan},
and collecting cut diagrams of diffractive type, we obtain for 
$2\chi_{a(j)}^{{\rm diffr}}$ 
the recursive equation shown in Fig.~\ref{fan-difr}:%
\begin{figure}[t]
\begin{center}\includegraphics[%
  width=14cm,
  height=4cm]{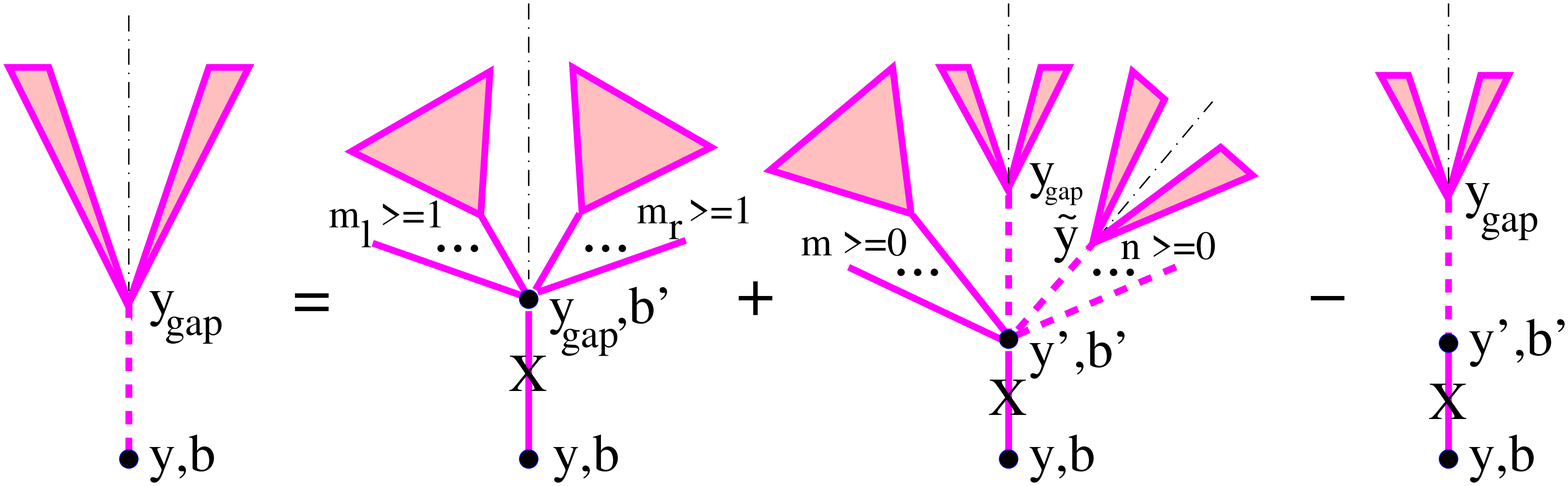}\end{center}
\vspace{-4mm}

\caption{Recursive equation for the diffractive {}``fan'' contribution
 $\chi_{a(j)}^{{\rm diffr}}(y,b,y_{{\rm gap}})$;
$y$, $b$ are rapidity and impact parameter distances between hadron
$a$ and the vertex in the {}``handle'' of the {}``fan'', $y_{{\rm gap}}$
is the size of the rapidity gap. Dot-dashed lines indicate the position
of the cut plane; cut pomerons are marked by crosses.\label{fan-difr}}
\end{figure}
\begin{eqnarray}
2\,\chi_{a(j)}^{{\rm diffr}}(y,b,y_{{\rm gap}})=
\frac{G}{\lambda_{a(j)}}\int\!\! d^{2}b'\;
\left\{ \left[1-e^{-\lambda_{a(j)}\,\chi_{a(j)}^{{\rm fan}}(y_{{\rm gap}},b')}
\right]^{2}\; G_{\mathbb{PP}}^{\mathbb{P}}(y-y_{{\rm gap}},|\vec{b}-\vec{b}'|)\right.\nonumber \\
+\int_{y_{{\rm gap}}}^{y}\!\! dy'\;2\lambda_{a(j)}\,
\chi_{a(j)}^{{\rm diffr}}(y',b',y_{{\rm gap}})\; G_{\mathbb{PP}}^{\mathbb{P}}(y-y',|\vec{b}-\vec{b}'|)\nonumber \\
\left.\times\left[\exp\!\left(-2\lambda_{a(j)}\,\chi_{a(j)}^{{\rm fan}}(y',b')+
\int_{y_{{\rm gap}}}^{y'}\!\! d\tilde{y}\;
2\lambda_{a(j)}\,\chi_{a(j)}^{{\rm diffr}}(y',b',\tilde{y})\right)-1\right]
\right\} .\label{ffan-difr}\end{eqnarray}
The first graph in the r.h.s.~of Fig.~\ref{fan-difr} is obtained
when the cut plane passes between the {}``fans'' connected to
the vertex $(y',b')$ in Fig.~\ref{ffan} (in which case we have
$y'=y_{{\rm gap}}$), with any number but at least one {}``fan''
remained on either side of the cut. Correspondingly the 2nd diagram
appears when the cut goes through at least one of the {}``fans'',
producing a rapidity gap of size $y_{{\rm gap}}$ inside. Then, the
vertex $(y',b')$ is coupled to the diffractive "fan" 
$\chi_{a(j)}^{{\rm diffr}}(y',b',y_{{\rm gap}})$
and to any number $m\geq0$ of uncut {}``fans'',
each of which may be positioned on either side of the cut. Also, any
number $n\geq0$ of additional diffractively cut {}``fans'' may
be connected to this vertex, provided all of them produce rapidity
gaps larger than $y_{{\rm gap}}$: $\tilde{y}_{i}\geq y_{{\rm gap}}$,
$i=1,...,n$. Finally, the last graph in the r.h.s.~of Fig.~\ref{fan-difr}
is to subtract the pomeron self-coupling contribution ($m=n=0$).

In turn, diffractive PDFs 
$x\,x_{\mathbb{P}}\,f_{I/a}^{{\rm diffr}}(x,x_{\mathbb{P}},Q_{0}^{2})$
are obtained from diagrams of Fig.~\ref{fan-difr} with $y=-\ln x$,
$y_{{\rm gap}}=-\ln x_{\mathbb{P}}$, replacing the down-most vertex
by pomeron-parton coupling (replacing the eikonal $G_{\mathbb{PP}}^{\mathbb{P}}$
in (\ref{ffan-difr}) by $G_{\mathbb{P}I}^{\mathbb{P}_{{\rm soft}}}$),
averaging over diffractive eigenstates of hadron $a$, and integrating
over impact parameter $b$:
\begin{eqnarray}
x\,x_{\mathbb{P}}\, f_{I/a}^{{\rm diffr}}(x,x_{\mathbb{P}},Q_{0}^{2})
=4\pi G\gamma_{\mathbb{P}}\, x^{-\Delta}\sum_{j}C_{a(j)}\int\!\! d^{2}b'\;
\left\{ \frac{1}{2}\gamma_{I}\!\left(\frac{x}{x_{\mathbb{P}}}\right)\, x_{\mathbb{P}}^{\Delta}\left[1-e^{-\lambda_{a(j)}\,\chi_{a(j)}^{{\rm fan}}(-\ln x_{\mathbb{P}},b')}\right]^{2}\right.\nonumber \\
+\int_{-\ln x_{\mathbb{P}}}^{-\ln x}\!\! dy'\;
\gamma_{I}\!\left(x\, e^{y'}\right)\, e^{-\Delta y'}\;
\lambda_{a(j)}\,\chi_{a(j)}^{{\rm diffr}}(y',b',-\ln x_{\mathbb{P}})\nonumber \\
\left.\times\left[\exp\!\left(-2\lambda_{a(j)}\,\chi_{a(j)}^{{\rm fan}}(y',b')+
\int_{-\ln x_{\mathbb{P}}}^{y'}\!\! d\tilde{y}\;
2\lambda_{a(j)}\,\chi_{a(j)}^{{\rm diffr}}(y',b',\tilde{y})\right)-1\right]\right\} .\label{pdf-difr}\end{eqnarray}

At arbitrary scale $Q^{2}$ we thus have\begin{eqnarray}
f_{I/a}^{{\rm diffr}}(x,x_{\mathbb{P}},Q^{2})=\sum_{J=g,q_{s}}\!\int_{x/x_{\mathbb{P}}}^{1}\!\!
\frac{dz}{z}\: E_{J\rightarrow I}^{{\rm QCD}}(z,Q_{0}^{2},Q^{2})\; 
f_{J/a}^{{\rm diffr}}(x/z,x_{\mathbb{P}},Q_{0}^{2})\,.\label{pdf-difr-qq}\end{eqnarray}
It is noteworthy that (\ref{pdf-difr}-\ref{pdf-difr-qq}) are only
applicable for high mass diffraction ($\beta=x/x_{\mathbb{P}}\ll1$),
as at moderate $\beta$ dominant contribution comes from the so-called
$q\bar{q}$ diffraction component \cite{nik92}, which is neglected
here.

\section{Results and discussion\label{sec:Results-and-discussion}}

The obtained formulas have been applied to calculate total, elastic,
and single diffraction proton-proton cross sections, elastic scattering slope
 $B_{pp}^{{\rm el}}(s)$,
as well as proton inclusive and diffractive SFs $F_{2/p}(x,Q^{2})$,
$F_{2/p}^{D(3)}(x,x_{\mathbb{P}},Q^{2})$. The latter are given to
leading order as\begin{eqnarray}
F_{2/p}(x,Q^{2})=\sum_{I=q,\bar{q}}e_{I}^{2}\, x\, f_{I/p}^{{\rm scr}}(x,Q^{2})+
F_{2/p}^{(c)}(x,Q^{2})\label{f2}\\
F_{2/p}^{D(3)}(x,x_{\mathbb{P}},Q^{2})=\sum_{I=q,\bar{q}}e_{I}^{2}\, x\, 
f_{I/p}^{{\rm diffr}}(x,x_{\mathbb{P}},Q^{2})\,.\label{f2d3}\end{eqnarray}
Here we use $f_{I/p}^{{\rm scr}}(x,Q^{2})=\int\! d^{2}b\;\tilde{f}_{I/p}^{{\rm scr}}(x,b,Q^{2})$,
$\tilde{f}_{I/p}^{{\rm scr}}(x,b,Q^{2})$ being defined by (\ref{pdf}) with
 $\tilde{f}_{J/a}^{{\rm scr}}(x,b,Q_{0}^{2})$  (see (\ref{pdf-scr}))
as the initial conditions for sea quarks and gluons; $f_{I/p}^{{\rm diffr}}(x,x_{\mathbb{P}},Q^{2})$
are given in (\ref{pdf-difr}-\ref{pdf-difr-qq}). The charm quark
contribution $F_{2/p}^{(c)}(x,Q^{2})$ has been calculated via the
photon-gluon fusion process \cite{grs}, using $m_{c}=1.3$ GeV for
the charm quark mass, and neglected in the diffractive structure function.
Single diffraction proton-proton cross section has been calculated as a sum
of the low and high mass diffraction contributions, 
$\sigma_{pp}^{{\rm SD}}(s)=2\,\sigma_{pp}^{{\rm LMD(targ)}}(s)
+2\,\sigma_{pp}^{{\rm HMD(targ)}}(s,y_{{\rm gap}})$,
the two latter being defined in (\ref{LMD-proj}), (\ref{sig-hmd}) correspondingly.
To compare with experimental data, the size of the rapidity gap for
the high mass diffraction has been determined from the condition that
the quasi-elastically scattered proton looses less than 5\% of its energy, 
$y_{{\rm gap}}=-\ln0.05$.

\begin{figure}[htb]
\begin{center}\includegraphics[%
  width=7cm,
  height=6.5cm]{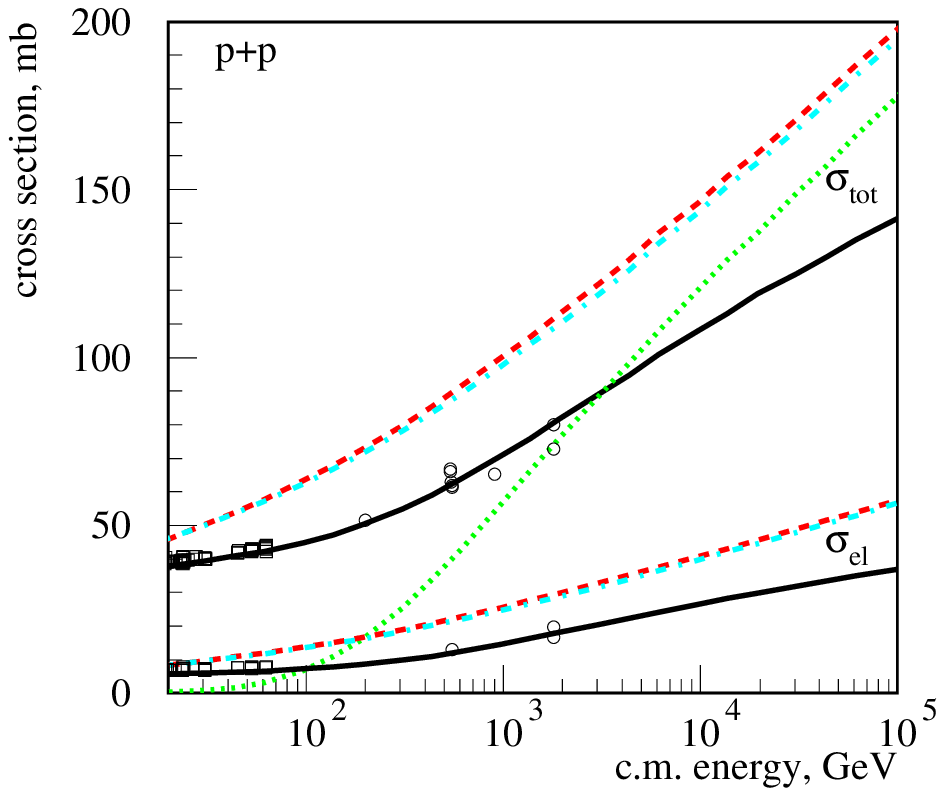}\hfill{}\includegraphics[%
  width=7cm,
  height=6.5cm]{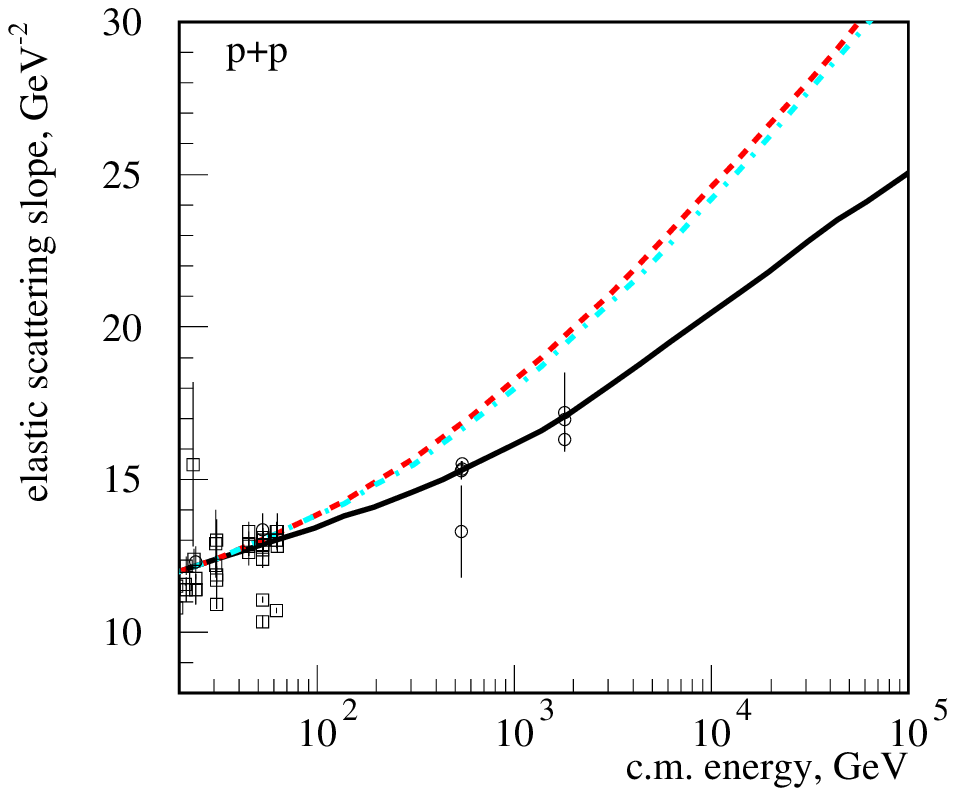}\end{center}
\vspace{-4mm}

\caption{Total and elastic proton-proton cross sections (left) and elastic
scattering slope (right) as calculated with (solid lines) and without
(dashed and dot-dashed (neglecting valence quark input) lines) enhanced
diagram contributions. Dotted line corresponds to $\sigma_{pp}^{{\rm tot}}(s)$,
calculated using only the factorized contribution of semi-hard processes
$\chi_{pp}^{{\rm sh(fact)}}(s,b)$, as explained in the text. The
compilation of data is from \cite{cas98}.\label{sig-pp}}
\end{figure}

\begin{figure}[htb]
\begin{center}\includegraphics[%
  width=14cm,
  height=10cm]{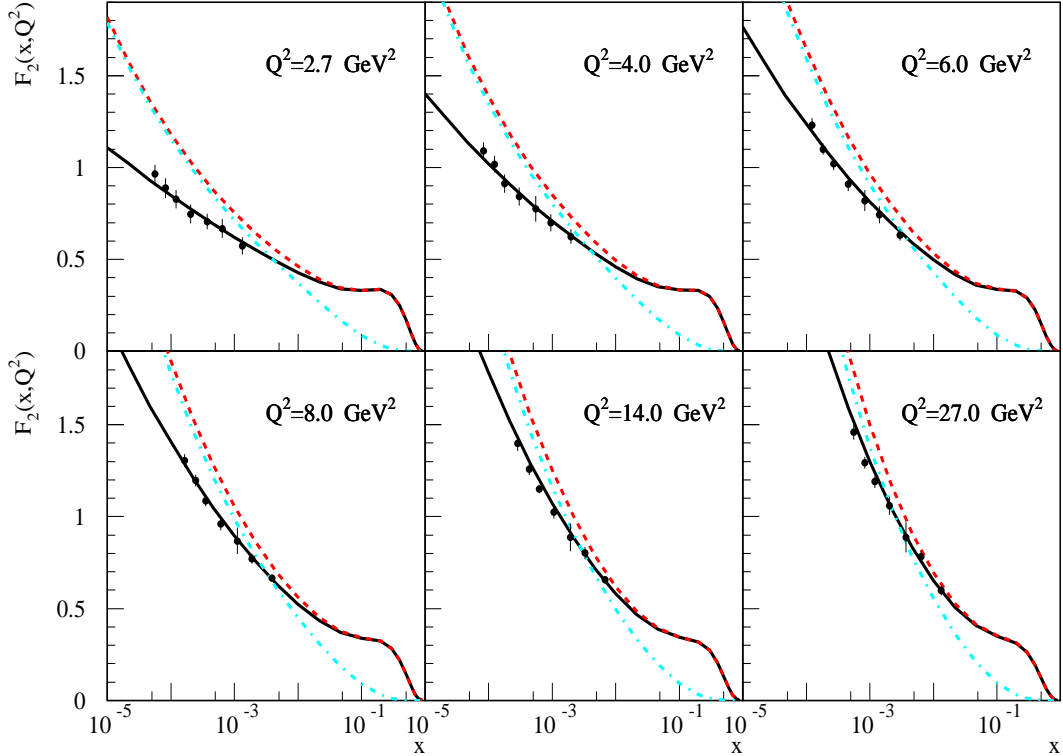}\end{center}
\vspace{-4mm}

\caption{Proton SF $F_{2/p}(x,Q^{2})$ calculated with (solid lines) and without
(dashed and dot-dashed (neglecting valence quark input) lines) enhanced
graph corrections, compared to data of the ZEUS forward plug calorimeter
\cite{che05}.\label{f2comp}}
\end{figure}

Concerning the parameter choice, we used  two-component diffraction scheme
with one {}``passive'' component, $\lambda_{p(2)}=0$, and with
the standard value of the shower enhancement coefficient
 $\lambda_{p(1)}=1/C_{p(1)}=\sqrt{1.5}$ \cite{kai82}.
It turned out that a reasonable agreement with data could be achieved
even for a rather low virtuality cutoff $Q_{0}^{2}=1$ GeV$^{2}$
for semi-hard processes; for the other parameters we obtained 
$\alpha_{{\rm P}}(0)=1.15$,
$\alpha_{{\rm P}}'(0)=0.075$ GeV$^{-2}$, $\gamma_{p}=5.6$ GeV$^{-1}$,
$R_{p}^{2}=2.15$ GeV$^{-2}$, $\gamma_{\mathbb{P}}=0.5$ GeV$^{-1}$,
$G=0.18$ GeV$^{2}$, $\beta_{g}=1$, $w_{qg}=0.22$. The results
for  $\sigma_{pp}^{{\rm tot}}$, $\sigma_{pp}^{{\rm el}}$,
$B_{pp}^{{\rm el}}$, and  $F_{2/p}$  are
plotted in Figs.~\ref{sig-pp},~\ref{f2comp}.%
 For comparison we show also the same quantities, calculated without enhanced diagram
contributions, i.e.~using the eikonal $\chi_{pp}(s,b)$, given in
(\ref{chi-total}), and the PDFs $\tilde f_{I/p}(x,b,Q^{2})$, defined by  (\ref{pdf}),
(\ref{sf-q0}-\ref{sf-val}). It is noteworthy that our analysis,
being devoted to high energy behavior of hadronic cross sections and
to the low $x$ asymptotics of SFs, is rather insensitive to  input
 PDFs of valence quarks $q_{v}(x,Q_{0}^{2})$ (see (\ref{sf-val})). For the
illustration, we repeated  the latter calculation neglecting the input valence
quark distribution, i.e.~setting $q_{v}(x,Q_{0}^{2})\equiv0$; the results are
shown in Figs.~\ref{sig-pp},~\ref{f2comp} by dot-dashed
lines.  As is easy to see, the obtained variations are very moderate in the
range of interest.  

\begin{figure}[htb]
\begin{center}\includegraphics[%
  width=14cm,
  height=12cm]{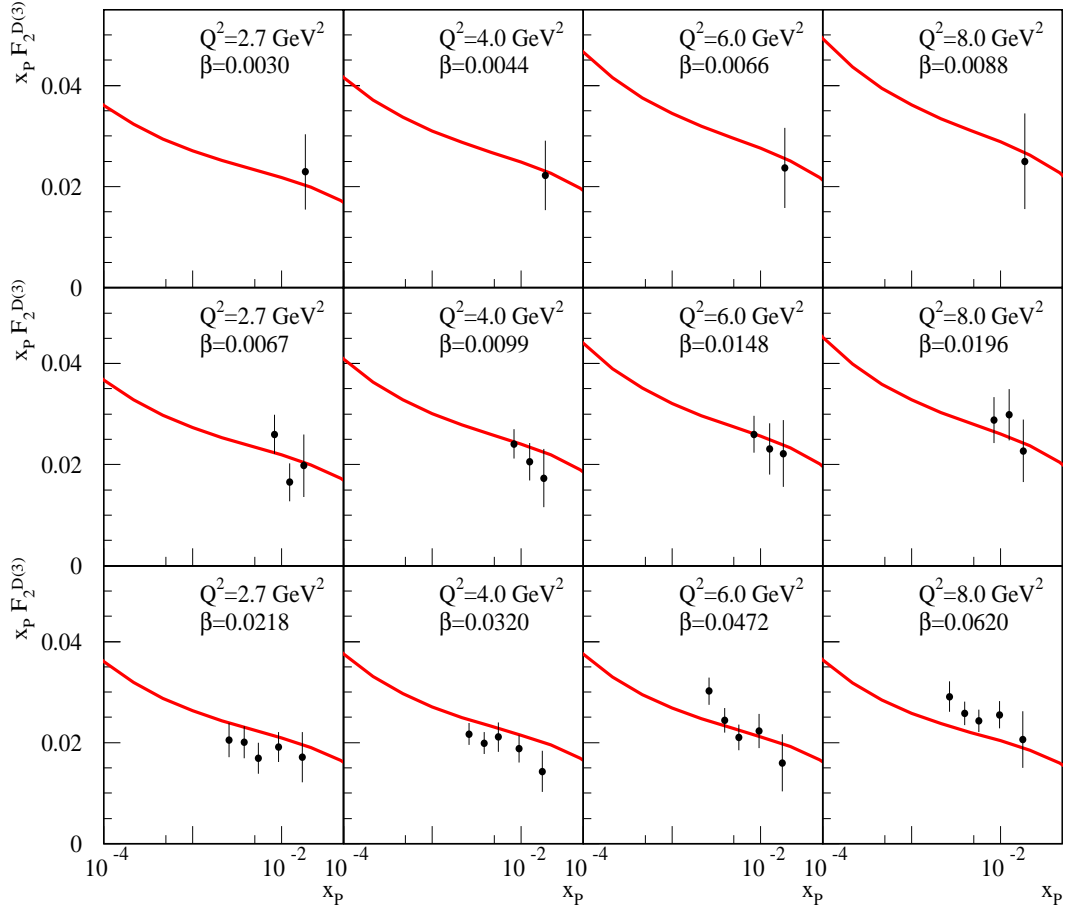}\end{center}
\vspace{-4mm}

\caption{Proton diffractive SF 
$x_{\mathbb{P}}\,F_{2/p}^{D(3)}(x,x_{\mathbb{P}},Q^{2})$,
compared to data of the ZEUS forward plug calorimeter \cite{che05}.\label{f2difr-comp}}
\end{figure}

\begin{figure}[htb]
\begin{center}\includegraphics[%
  width=8cm,
  height=6.5cm]{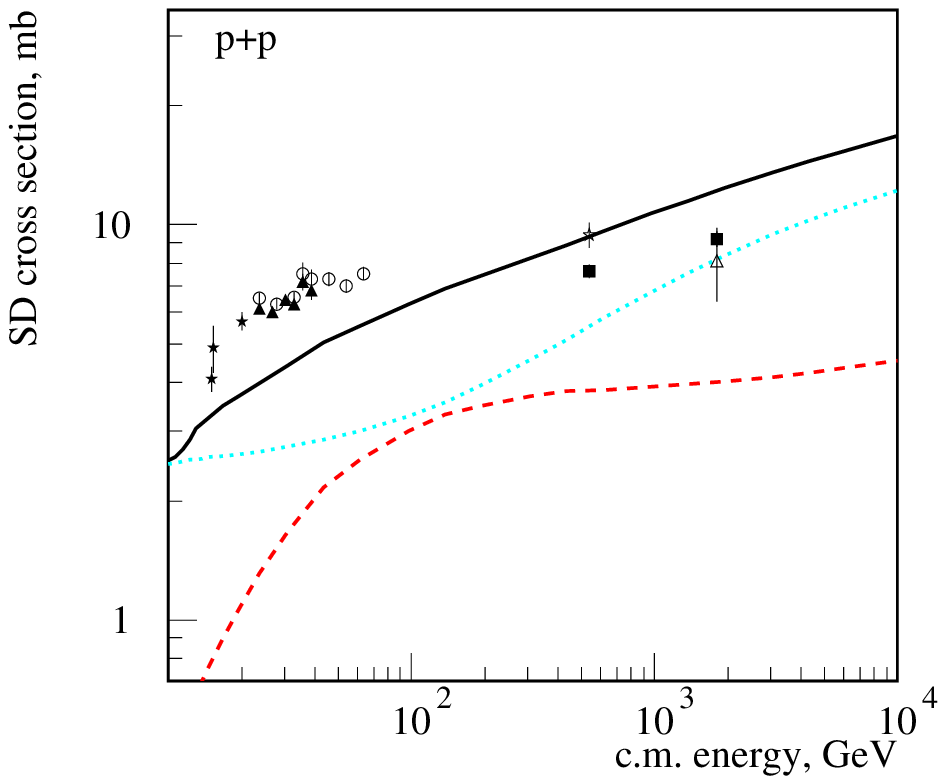}\end{center}
\vspace{-4mm}

\caption{Calculated total, high and low mass single diffraction proton-proton 
cross sections -- solid, dashed, and dotted lines correspondingly. The
compilation of data is from \cite{gou04}.\label{sd-fig} }
\end{figure}

In Figs.~\ref{f2difr-comp},~\ref{sd-fig}%
 the calculated proton diffractive SF $F_{2/p}^{D(3)}$ and single diffraction proton-proton
 cross section $\sigma_{pp}^{{\rm SD}}$ are compared to experimental data; the partial 
 contributions of low and high mass diffraction are also shown in Fig.~\ref{sd-fig}.
 In general, a satisfactory agreement with measurements is observed both for the diffractive
 DIS contribution and for the ``soft'' hadronic diffraction. The obtained energy rise of 
 $\sigma_{pp}^{{\rm SD}}(s)$  is somewhat steeper than observed experimentally, due to the 
 rapid increase of the low  mass diffraction contribution, as seen  in Fig.~\ref{sd-fig}.
 This is a consequence of using the simple ``passive'' component (quasi-eikonal) approach,
 which leads to a proportionality between $\sigma_{ad}^{{\rm el}}$ and 
 $\sigma_{ad}^{{\rm LMD}}$, 
 i.e.~$\sigma_{ad}^{{\rm LMD}}(s)/\sigma_{ad}^{{\rm el}}(s)\equiv {\rm const}$,
 as one can see from (\ref{sigma-el}), (\ref{LMD-proj}). Employing a general multi-component
 scheme with more than one ``active'' component,
  one can substantially reduce the energy dependence of the low  mass diffraction contribution
 and improve the agreement with data.
It is noteworthy that the obtained moderate energy rise of the high mass diffraction component
is not only due to the usual suppression of rapidity gap topologies by the
elastic form-factor, but also due to the unitarization of the bare
contribution of diffractively cut graphs, which originates from
additional re-scattering processes on both projectile and target hadrons;
each pomeron, connected to the cut multi-pomeron vertex at the edge of the
gap, appears to be a ``handle'' of
 either cut or uncut {}``net fan'' sub-graph, as can be seen
in Fig.~\ref{1-gap-fig}.

Let us now note the differences with our previous treatment \cite{ost06},
which used only soft pomeron contributions and was based on the assumption
of $\pi$-meson dominance of multi-pomeron vertices. Here, considering
contributions of both soft and semi-hard processes and assuming a
small slope of multi-pomeron vertices, we obtained an unusually small
value for the soft pomeron slope. This is because the enhanced diagram
contribution $\chi_{ad(jk)}^{{\rm enh}}(s,b)$, defined in (\ref{enh-tot}),
is most significant in the region of comparatively small impact parameters,
being characterized by a somewhat smaller effective slope than the one
of the soft pomeron. On the other hand, the obtained values of the
pomeron intercept $\alpha_{{\rm P}}(0)=1.15$ and of the triple-pomeron
coupling $r_{3\mathbb{P}}=4\pi\, G\,\gamma_{\mathbb{P}}^{3}\simeq0.28$
GeV$^{-1}$ are not too different from the
 ones in \cite{ost06}: 1.18 and 0.18
GeV$^{-1}$ correspondingly.

We would like to stress also an important feature of the presented approach:
  the full interaction eikonal (\ref{chi_ad-tot}),
which includes the enhanced diagram contribution (\ref{enh-tot}),
can no longer be expressed in the usual factorized form 
 (\ref{chi-mini-jet}),~(\ref{chi-total}).
 In particular, non-linear screening corrections
to the contribution of semi-hard processes can not be simply absorbed
into the re-defined PDFs $\tilde{f}_{I/a}^{{\rm scr}}(x,b,Q^{2})$. Significant
non-factorizable corrections come from graphs where at least one pomeron
is exchanged in parallel to the hardest parton scattering process,
with the simplest example given by the 1st diagram in the r.h.s.~of
Fig.~\ref{3p-vertex}. In fact, such contributions play an important
role for reaching the consistency between total hadronic cross sections
and structure functions. For the illustration, in Fig.~\ref{sig-pp}
shown also the result for $\sigma_{pp}^{{\rm tot}}(s)$, as calculated
using only the factorized semi-hard contribution 
$\chi_{pp}^{{\rm sh(fact)}}(s,b)$,
i.e.~using the eikonal (\ref{chi-total}) with $\chi_{pp}^{\mathbb{P}_{{\rm soft}}}(s,b)\equiv0$
and with the PDFs $\tilde{f}_{I/p}(x,b,Q^{2})$ being replaced by
 $\tilde{f}_{I/p}^{{\rm scr}}(x,b,Q^{2})$.
It is easy to see that in such a case the cross section rises with
energy much faster than obtained before with the full eikonal (\ref{chi_ad-tot}),
even though the contribution of soft processes is neglected.

\begin{figure}[htb]
\begin{center}\includegraphics[%
  width=10cm,
  height=4cm]{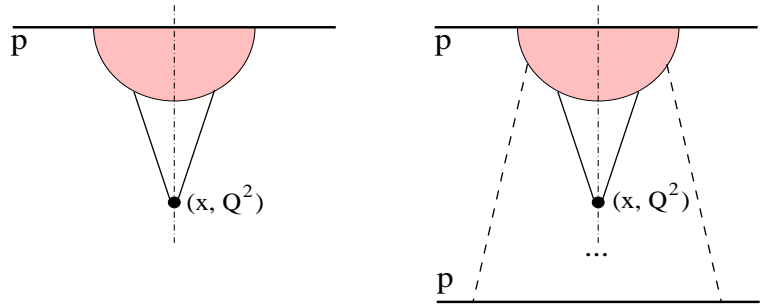}\end{center}
\vspace{-4mm}

\caption{Schematic view of parton distributions as \char`\"{}seen\char`\"{}
in DIS (left) and in proton-proton collision (right). Low $x$ parton (sea quark
or gluon) originates from the initial state {}``blob'' and interacts
with a highly virtual {}``probe''. In proton-proton interaction
 the initial {}``blob'' itself is affected by the collision
process -- due to additional soft re-scatterings on the target, indicated
by dashed lines.\label{pdf-non} }
\end{figure}

To additionally clarify this point, let us consider PDFs in the low
$x$ limit, sketched in Fig.~\ref{pdf-non},%
 as {}``seen'' in DIS reactions and in hadronic collisions. In the
former case, depicted on the left, all non-linear corrections to parton
dynamics come from re-scattering on constituent partons of the same
parent hadron, being hidden in the upper {}``blob'' in the Figure.
The corresponding PDFs are thus described by {}``fan'' diagram contributions.
On the other hand, in hadron-hadron interaction one encounters additional
re-scatterings on constituent partons of the partner hadron, indicated
symbolically in the r.h.s.~graph of Fig.~\ref{pdf-non} as dashed
lines connecting the upper {}``blob'' with the target hadron. Parton
cascades, which mediate these additional re-scattering processes,
may couple both to independent constituents of the projectile hadron,
which would lead to the usual multiple scattering picture of Fig.~\ref{multiple},
or to {}``soft'' parents of the given high-$p_{t}$ parton, as shown
in Fig.~\ref{ladder-merge}(right), thus modifying the initial state
parton evolution. In the high energy asymptotics the second configuration
dominates, being enhanced by logarithmic factors. As a consequence,
both the interaction eikonal and correspondingly the cross sections
for particular inelastic final states can not be expressed via universal
PDFs of free hadrons. In principle, this is not surprising, keeping
in mind that QCD collinear factorization holds only for fully inclusive
quantities \cite{col89}. In the present approach such initial state {}``blobs'',
with the re-scatterings included, are described by the {}``net fan''
contributions. The latter may be regarded as a kind of reaction-dependent
{}``parton distributions'', which are probed during interaction
and are thus affected by the surrounding medium.

\begin{figure}[htb]
\begin{center}\includegraphics[%
  width=2.5cm,
  height=4cm]{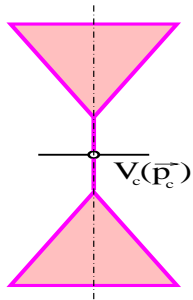}\end{center}
\vspace{-4mm}

\caption{Diagrams contributing to single inclusive cross sections; $V_{c}(\vec{p}_{c})$
is the particle $c$ emission vertex from a cut pomeron. \label{inclus} }
\end{figure}

At the same moment, due to the AGK cancellations \cite{agk}, the
above-discussed non-factorizable graphs give negligible contribution
to inclusive high-$p_{t}$ jet cross sections. Single inclusive particle
spectra are defined as usual by the diagrams of
 Fig.~\ref{inclus} \cite{mue81},%
 as far as higher twist effects due to final high-$p_{t}$ parton
re-scattering are neglected \cite{qiu04}. In particular, inclusive jet
cross sections are thus given in the usual factorized form \cite{owe87}:
as the convolution of hadronic PDFs $f_{I(J)/a(d)}^{{\rm scr}}(x^{+(-)},Q^{2})$
and matrix elements for parton emission.

It is noteworthy that the presented results have been obtained under
the assumption on the
eikonal structure (\ref{g_mn}) of multi-pomeron vertices. In principle,
one may restrict himself with only triple-pomeron vertices, i.e. set
$\gamma_{\mathbb{P}}=0$ in (\ref{g_mn}). In practical terms this
would mean to replace the constant $G$ by
 $r_{3\mathbb{P}}/(4\pi\,\gamma_{\mathbb{P}}^{3})$
and to consider the limit $\gamma_{\mathbb{P}}\rightarrow0$ in all
the obtained formulas. However, in such a case the scheme would be
incomplete: one will need to include also the contributions of pomeron
{}``loop'' diagrams, which contain internal multi-pomeron vertices connected
to each other by at least two pomerons. In the eikonal scheme these
contributions are suppressed by exponential factors \cite{car74,ost06},
which allowed to neglect them in the present analysis.

It is also worth reminding that throughout this work we neglected the
effects of pomeron-pomeron coupling in high ($|q|^{2}>Q_{0}^{2}$)
virtuality region. One may expect that in very high energy asymptotics
those contributions also become significant. However, being suppressed
as $1/q^{4}$, they should manifest themselves only in the sufficiently
{}``black'' region of moderately small impact parameters, where
parton densities are high enough to compensate the mentioned suppression.
Therefore, we do not expect a significant modification of the cross
section results obtained, when such contributions are taken into account.

In conclusion, accounting for non-linear screening effects, one can
obtain a consistent description of hadronic cross sections and of
corresponding structure functions, using a fixed energy-independent
virtuality cutoff for the contribution of semi-hard processes. 
On the other hand, a general  consistency is observed between {}``soft''
hadronic diffraction and the one measured in DIS processes. An important
feature of the proposed scheme is that the contribution of semi-hard
processes to the interaction eikonal contains a significant non-factorizable
part. This circumstance has to be taken into account if one attempts
to extract information on parton saturation from the behavior of hadronic
cross sections. On the other hand, by virtue of the AGK cancellations
the corresponding diagrams do not contribute to inclusive parton jet
spectra and the scheme preserves the QCD factorization picture.

\section*{Appendix}

\begin{figure}[htb]
\begin{center}\includegraphics[%
  width=14cm,
  height=4cm]{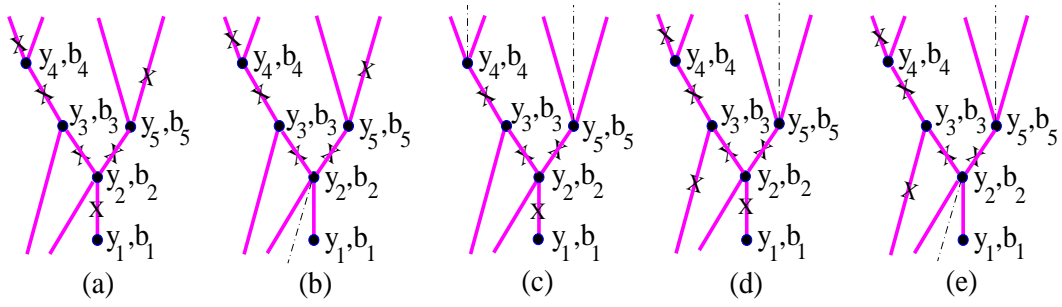}\end{center}
\vspace{-4mm}

\caption{Examples of graphs obtained by cutting the same projectile {}``net
fan'' diagram: in the graphs (a), (b), (c) we have a {}``fan''-like
structure of cut pomerons (marked by crosses); in the diagrams (d),
(e) the cut pomeron exchanged between the vertex ($y_{3},b_{3}$)
and the target forms a {}``zig-zag'' with the {}``handle'' of
the {}``fan''. The cut plane is indicated by dot-dashed lines.\label{cut examples}}
\end{figure}

We are going to derive contributions of diffractive cuts of general
enhanced graphs of Fig.~\ref{enh-full}. It is convenient to start
from the analysis of unitarity cuts of {}``net fan'' diagrams of
Fig.~\ref{freve}. One can separate them in two classes: in the first
sub-set cut pomerons form a {}``fan''-like structure, some examples
shown in Fig.~\ref{cut examples}~(a),~(b),~(c);%
 in the diagrams of the second kind some intermediate vertices contain
\textit{cut} pomerons connected to the partner hadron $d$, see Fig.~\ref{cut examples}~(d),~(e),
such that these pomerons are arranged in a {}``zig-zag'' way with
respect to the {}``handle'' of the {}``fan''.

\begin{figure}[t]
\begin{center}\includegraphics[%
  width=15cm,
  height=4cm]{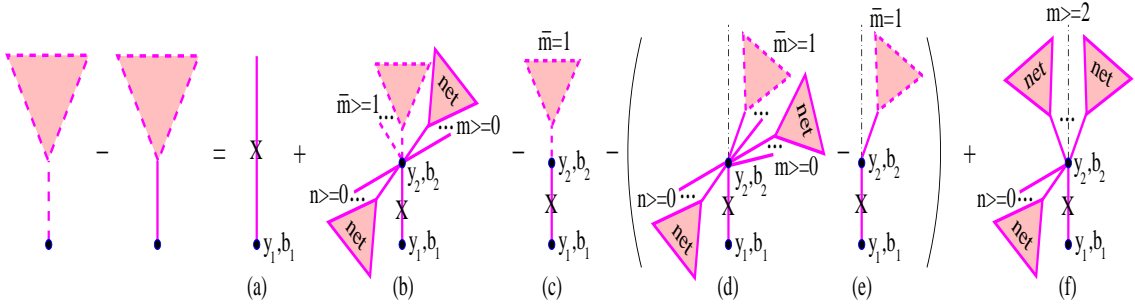}\end{center}
\vspace{-4mm}

\caption{Recursive equation for the contribution $2\bar{\chi}_{a(j)|d(k)}^{{\rm fan}}-2\tilde{\chi}_{a(j)|d(k)}^{{\rm fan}}$
of {}``fan''-like cuts of {}``net-fan'' diagrams, where the cut
plane goes through the {}``handle'' of the {}``fan''. Cut pomerons
are marked by crosses, the cut plane is indicated by dot-dashed lines.\label{fan-cut-fig}}
\end{figure}

\begin{figure}[t]
\begin{center}\includegraphics[%
  width=15cm,
  height=4cm]{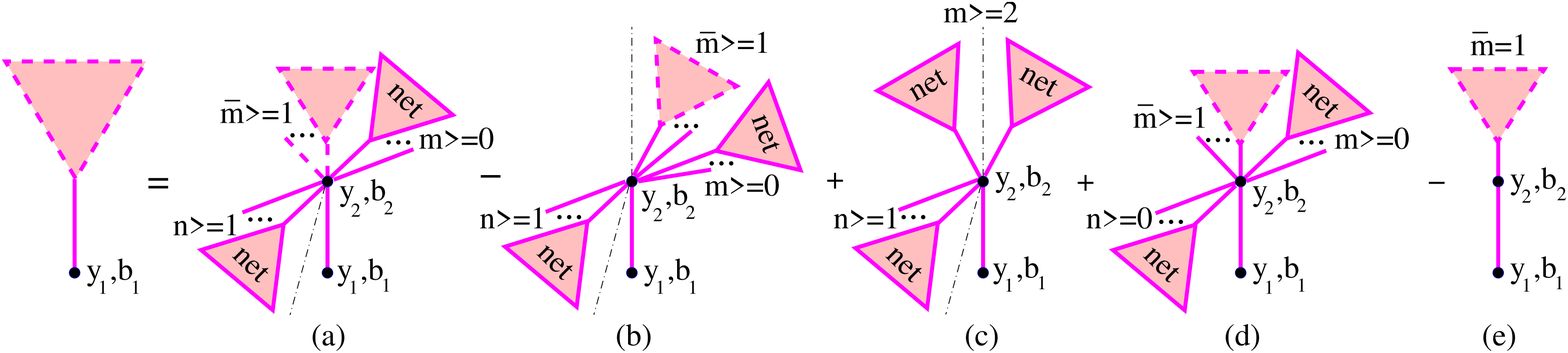}\end{center}
\vspace{-4mm}

\caption{Recursive equation for the contribution $2\tilde{\chi}_{a(j)|d(k)}^{{\rm fan}}$
of {}``fan''-like cuts of {}``net-fan'' diagrams, where the {}``handle''
of the {}``fan'' remains uncut. Cut pomerons are marked by crosses,
the cut plane is indicated by dot-dashed lines.\label{fan-hole-fig}}
\end{figure}

Let us consider the first class and obtain separately both the total
contribution of ``fan''-like cuts $2\bar{\chi}_{a(j)|d(k)}^{{\rm fan}}$
and a part of it, formed by diagrams with the {}``handle'' of the
{}``fan'' being uncut (see Fig.~\ref{cut examples}~(b)) - $2\tilde{\chi}_{a(j)|d(k)}^{{\rm fan}}$.
Applying AGK cutting rules \cite{agk} to the general {}``net fan''
graphs of Fig.~\ref{freve} and collecting contributions of cuts
of desirable structures we obtain for $2\bar{\chi}_{a(j)|d(k)}^{{\rm fan}}-2\tilde{\chi}_{a(j)|d(k)}^{{\rm fan}}$,
$2\tilde{\chi}_{a(j)|d(k)}^{{\rm fan}}$ the representations of Figs.~\ref{fan-cut-fig},
\ref{fan-hole-fig},%
 which gives\begin{eqnarray}
2\,\bar{\chi}_{a(j)|d(k)}^{{\rm fan}}(y_{1},\vec{b}_{1}|Y,\vec{b})-2\,\tilde{\chi}_{a(j)|d(k)}^{{\rm fan}}(y_{1},\vec{b}_{1}|Y,\vec{b})=2\,\chi_{a\mathbb{P}}^{\mathbb{P}}(y_{1},b_{1})+\frac{G}{\lambda_{a(j)}}\int_{0}^{y_{1}}\!\! dy_{2}\int\!\! d^{2}b_{2}\nonumber \\
\times G_{\mathbb{PP}}^{\mathbb{P}}(y_{1}-y_{2},|\vec{b}_{1}-\vec{b}_{2}|)\;\left\{ \left(e^{2\lambda_{a(j)}\,\bar{\chi}_{a(j)|d(k)}^{{\rm fan}}}-1\right)e^{-2\lambda_{a(j)}\,\chi_{a(j)|d(k)}^{{\rm net}}-2\lambda_{d(k)}\,\chi_{d(k)|a(j)}^{{\rm net}}}\right.\nonumber \\
-2\lambda_{a(j)}\,\bar{\chi}_{a(j)|d(k)}^{{\rm fan}}-2\left[\left(e^{\lambda_{a(j)}\,\tilde{\chi}_{a(j)|d(k)}^{{\rm fan}}}-1\right)e^{-\lambda_{a(j)}\,\chi_{a(j)|d(k)}^{{\rm net}}-2\lambda_{d(k)}\,\chi_{d(k)|a(j)}^{{\rm net}}}-\lambda_{a(j)}\,\tilde{\chi}_{a(j)|d(k)}^{{\rm fan}}\right]\nonumber \\
\left.+\left(1-e^{-\lambda_{a(j)}\,\chi_{a(j)|d(k)}^{{\rm net}}}\right)^{2}\, e^{-2\lambda_{d(k)}\,\chi_{d(k)|a(j)}^{{\rm net}}}\right\} \label{chi-cut-fan}\\
2\,\tilde{\chi}_{a(j)|d(k)}^{{\rm fan}}(y_{1},\vec{b}_{1}|Y,\vec{b})=\frac{G}{\lambda_{a(j)}}\int_{0}^{y_{1}}\!\! dy_{2}\int\!\! d^{2}b_{2}\; G_{\mathbb{PP}}^{\mathbb{P}}(y_{1}-y_{2},|\vec{b}_{1}-\vec{b}_{2}|)\nonumber \\
\times\left\{ \left(1-e^{-\lambda_{d(k)}\,\chi_{d(k)|a(j)}^{{\rm net}}}\right)e^{-\lambda_{d(k)}\,\chi_{d(k)|a(j)}^{{\rm net}}}\;\left[\left(e^{2\lambda_{a(j)}\,\bar{\chi}_{a(j)|d(k)}^{{\rm fan}}}-1\right)e^{-2\lambda_{a(j)}\,\chi_{a(j)|d(k)}^{{\rm net}}}\right.\right.\nonumber \\
\left.-2\left(e^{\lambda_{a(j)}\,\tilde{\chi}_{a(j)|d(k)}^{{\rm fan}}}-1\right)e^{-\lambda_{a(j)}\,\chi_{a(j)|d(k)}^{{\rm net}}}+\left(1-e^{-\lambda_{a(j)}\,\chi_{a(j)|d(k)}^{{\rm net}}}\right)^{2}\right]\nonumber \\
\left.+2\left[\left(e^{\lambda_{a(j)}\,\tilde{\chi}_{a(j)|d(k)}^{{\rm fan}}}-1\right)e^{-\lambda_{a(j)}\,\chi_{a(j)|d(k)}^{{\rm net}}-\lambda_{d(k)}\,\chi_{d(k)|a(j)}^{{\rm net}}}-\lambda_{a(j)}\,\tilde{\chi}_{a(j)|d(k)}^{{\rm fan}}\right]\right\} .\label{chi-cut-handle}\end{eqnarray}
Here the arguments of the eikonals in the r.h.s.~of (\ref{chi-cut-fan}),
(\ref{chi-cut-handle}) are understood as $X_{a(j)|d(k)}=X_{a(j)|d(k)}(y_{2},\vec{b}_{2}|Y,\vec{b})$,
$X_{d(k)|a(j)}=X_{d(k)|a(j)}(Y-y_{2},\vec{b}-\vec{b}_{2}|Y,\vec{b})$;
$X=\chi^{{\rm net}}$, $\bar{\chi}^{{\rm fan}}$, $\tilde{\chi}^{{\rm fan}}$.
The first diagram in the r.h.s.~of Fig.~\ref{fan-cut-fig} is obtained
cutting the single pomeron exchanged between hadron $a$ and the vertex
$(y_{1},b_{1})$ in the r.h.s.~of Fig.~\ref{freve}, whereas the
other diagrams emerge when the 2nd graph in the r.h.s.~of Fig.~\ref{freve}
is cut in such a way that all cut pomerons are arranged in a {}``fan''-like
structure and the cut plane passes through the {}``handle'' of the
{}``fan''. In graph (b) the vertex $(y_{2},b_{2})$ couples together
$\bar{m}\geq1$ cut projectile {}``net fans'', each one characterized
by a {}``fan''-like structure of cuts, and any numbers $m,n\geq0$
of uncut projectile and target {}``net fans''. Here one has to subtract
pomeron self-coupling contribution ($\bar{m}=1$; $m,n=0$) - graph
(c), as well as the contributions of graphs (d), (e), where in all
$\bar{m}$ cut {}``fans'', connected to the vertex $(y_{2},b_{2})$,
the {}``handles'' of the {}``fans'' remain uncut and all these
{}``handles'' and all the $m$ uncut projectile {}``net fans''
are situated on the same side of the cut plane. Finally, in graph
(f) the cut plane passes between $m\geq2$ uncut projectile {}``net
fans'',  with at least one remained on either side of the cut. In the
recursive representation of Fig.~\ref{fan-hole-fig} for the contribution
$2\tilde{\chi}_{a(j)|d(k)}^{{\rm fan}}$ the graphs (a), (b), (c)
in the r.h.s.~of the Figure are similar to the diagrams (b), (d),
(f) of Fig.~\ref{fan-cut-fig} correspondingly, with the difference
that the {}``handle'' of the {}``fan'' is now uncut. Therefore,
there are $n\geq1$ uncut target {}``net fans'' connected to the
vertex $(y_{2},b_{2})$, such that at least one of the latter is positioned
on the opposite side of the cut plane with respect to the {}``handle''
pomeron. On the other hand, one has to add graph (d), where the vertex
$(y_{2},b_{2})$ couples together $\bar{m}\geq1$ projectile {}``net
fans'', which are cut in a {}``fan''-like way and have their {}``handles''
uncut and positioned on the same side of the cut plane, together
with any numbers $m\geq0$ of projectile and $n\geq0$ of target uncut
{}``net fans'', such that the vertex $(y_{2},b_{2})$ remains uncut.
Here one has to subtract the pomeron self-coupling ($\bar{m}=1$;
$m,n=0$) - graph (e).

Adding (\ref{chi-cut-handle}) to (\ref{chi-cut-fan}), we obtain\begin{eqnarray}
2\,\bar{\chi}_{a(j)|d(k)}^{{\rm fan}}(y_{1},\vec{b}_{1}|Y,\vec{b})=2\,\chi_{a\mathbb{P}}^{\mathbb{P}}(y_{1},b_{1})+\frac{G}{\lambda_{a(j)}}\int_{0}^{y_{1}}\!\! dy_{2}\int\!\! d^{2}b_{2}\; G_{\mathbb{PP}}^{\mathbb{P}}(y_{1}-y_{2},|\vec{b}_{1}-\vec{b}_{2}|)\nonumber \\
\times\left\{ \left[\left(e^{2\lambda_{a(j)}\,\bar{\chi}_{a(j)|d(k)}^{{\rm fan}}}-1\right)e^{-2\lambda_{a(j)}\,\chi_{a(j)|d(k)}^{{\rm net}}}+\left(1-e^{-\lambda_{a(j)}\,\chi_{a(j)|d(k)}^{{\rm net}}}\right)^{2}\right]\, e^{-\lambda_{d(k)}\,\chi_{d(k)|a(j)}^{{\rm net}}}\right.\nonumber \\
\left.-2\lambda_{a(j)}\,\bar{\chi}_{a(j)|d(k)}^{{\rm fan}}\right\} .\label{chi-cut-fan1}\end{eqnarray}

Comparing with (\ref{net-fan}), we see that the solution of (\ref{chi-cut-fan1})
is\begin{eqnarray}
\bar{\chi}_{a(j)|d(k)}^{{\rm fan}}(y_{1},\vec{b}_{1}|Y,\vec{b})\equiv\chi_{a(j)|d(k)}^{{\rm net}}(y_{1},\vec{b}_{1}|Y,\vec{b}).\label{equiv}\end{eqnarray}

Correspondingly, using (\ref{equiv}), we can simplify (\ref{chi-cut-handle})
to obtain\begin{eqnarray}
\tilde{\chi}_{a(j)|d(k)}^{{\rm fan}}(y_{1},\vec{b}_{1}|Y,\vec{b})=\frac{G}{\lambda_{a(j)}}\int_{0}^{y_{1}}\!\! dy_{2}\int\!\! d^{2}b_{2}\; G_{\mathbb{PP}}^{\mathbb{P}}(y_{1}-y_{2},|\vec{b}_{1}-\vec{b}_{2}|)\,\left\{ \left(1-e^{-\lambda_{a(j)}\,\chi_{a(j)|d(k)}^{{\rm net}}}\right)\right.\nonumber \\
\left.\times e^{-\lambda_{d(k)}\,\chi_{d(k)|a(j)}^{{\rm net}}}-\left(1-e^{\lambda_{a(j)}\,\tilde{\chi}_{a(j)|d(k)}^{{\rm fan}}-\lambda_{a(j)}\,\chi_{a(j)|d(k)}^{{\rm net}}}\right)\, e^{-2\lambda_{d(k)}\,\chi_{d(k)|a(j)}^{{\rm net}}}-\lambda_{a(j)}\,\tilde{\chi}_{a(j)|d(k)}^{{\rm fan}}\right\} .\label{chi-cut-handle1}\end{eqnarray}

\begin{figure}[t]
\begin{center}\includegraphics[%
  width=6cm,
  height=4cm]{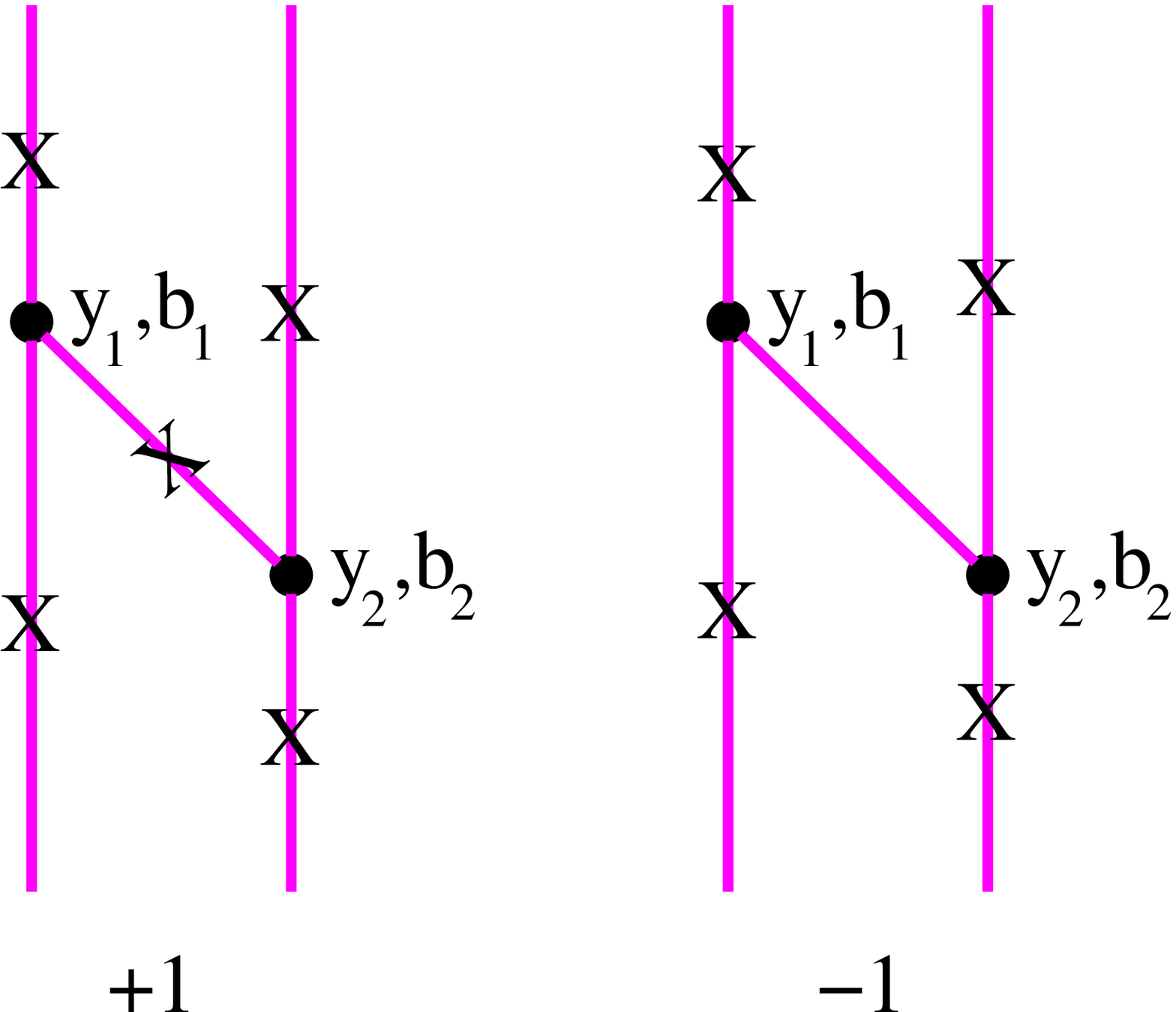}\end{center}
\vspace{-4mm}

\caption{Lowest order cut diagrams of {}``zig-zag'' type; cut pomerons are
marked by crosses. Numbers below the graphs indicate relative weights
of the corresponding contributions.\label{h-cut-fig}}
\end{figure}

As the summary contribution of all cuts of {}``net fan'' graphs
should be equal to twice the uncut one $2\chi_{a(j)|d(k)}^{{\rm net}}$
(the total discontinuity of an elastic scattering amplitude equals
to twice the imaginary part of the amplitude), from (\ref{equiv})
we can conclude that contributions of different {}``zig-zag'' cuts
of {}``net fan'' graphs (see the examples in Fig.~\ref{cut examples}~(d),~(e))
cancel each other, which can be also verified explicitely \cite{ost07}.
Moreover, it is possible to show that a similar cancellation takes
place for all unitarity cuts of the graphs of Fig.~\ref{enh-full},
which give rise to {}``zig-zag'' sub-structures formed by cut pomerons,
and that such cuts do not contribute to diffractive topologies \cite{ost07}.
As an illustration, let us compare the two diagrams in Fig.~\ref{h-cut-fig},%
 whose contributions are equal up to a sign and precisely cancel each
other. The right-hand graph provides a screening correction to the
eikonal configuration with two cut pomerons. On the other hand, the
left-hand graph introduces a new process, with the weight being equal
to the one of the screening correction above, and with the particle
production pattern being almost identical to the previous two cut
pomeron configuration; the only difference arises from the cut pomeron
exchanged between the vertices ($y_{1},b_{1}$) and ($y_{2},b_{2}$),
which leads to additional particle production in the rapidity interval
$[y_{1},y_{2}]$. However, this interval is already covered by particles,
which result from the left-most cut pomeron in the two graphs. Correspondingly,
the  rapidity gap structure of the event remains unchanged.

Thus, in the following we shall restrict ourselves to the analysis
of  ``tree''-like cuts of the diagrams of Fig.~\ref{enh-full},
whose contributions can be expressed via the ones of {}``fan''-like
cuts of {}``net fan'' graphs. Before we proceed further, let us
calculate the contributions of sub-samples of {}``fan''-like cuts
of {}``net fan'' graphs, which give rise to a rapidity gap of size
$\geq y_{{\rm gap}}$ between hadron $a$ and the nearest particle produced
after the gap, an example shown in Fig.~\ref{cut examples}~(c)
($y_{{\rm gap}}=y_{4}$). For simplicity, we shall use the two component
Good-Walker approach with one passive component, $\lambda_{a(2)}\equiv0$,
$C_{a(1)}\equiv1/\lambda_{a(1)}$, and neglect sub-dominant contributions
of diffractive cuts which leave the {}``handle'' of the {}``fan''
uncut (general derivation proceeds identically).

\begin{figure}[t]
\begin{center}\includegraphics[%
  width=13cm,
  height=4cm]{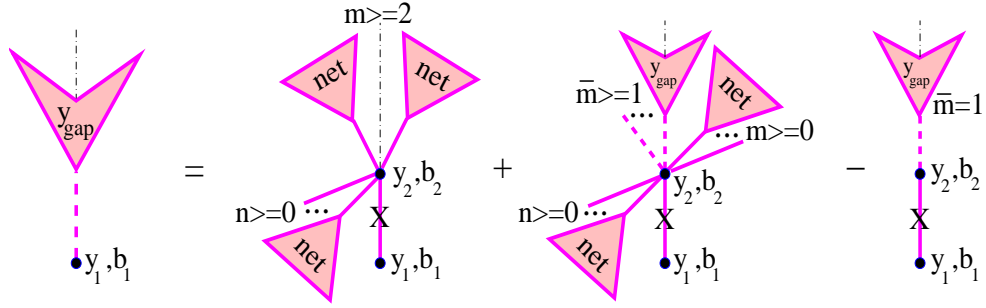}\end{center}
\vspace{-4mm}

\caption{Recursive equation for the contribution $2\bar{\chi}_{a(1)|d(1)}^{{\rm gap}}$
of {}``fan''-like diffractive cuts of {}``net-fan'' diagrams.
Cut pomerons are marked by crosses, the cut plane is indicated by
dot-dashed lines.\label{fan-difr-fig}}
\end{figure}

For the contribution of {}``fan''-like diffractive cuts $2\bar{\chi}_{a(1)|d(1)}^{{\rm gap}}$
we can easily obtain, similarly to Fig.~\ref{fan-difr}, the recursive
representation of Fig.~\ref{fan-difr-fig},%
 which gives\begin{eqnarray}
2\,\bar{\chi}_{a(1)|d(1)}^{{\rm gap}}(y_{1},\vec{b}_{1},y_{{\rm gap}}|Y,\vec{b})=
\frac{G}{\lambda_{a(1)}}\int_{y_{{\rm gap}}}^{y_{1}}\!\! dy_{2}\int\!\! d^{2}b_{2}\;
 G_{\mathbb{PP}}^{\mathbb{P}}(y_{1}-y_{2},|\vec{b}_{1}-\vec{b}_{2}|)\nonumber \\
\times\left\{ \left[\left(1-e^{-\lambda_{a(1)}\,
\chi_{a(1)|d(1)}^{{\rm net}}(y_{2},\vec{b}_{2}|Y,\vec{b})}\right)^{2}+
\left(e^{2\lambda_{a(1)}\,\bar{\chi}_{a(1)|d(1)}^{{\rm gap}}
(y_{2},\vec{b}_{2},y_{{\rm gap}}|Y,\vec{b})}-1\right) \right.\right.
\nonumber \\ \left.\left.\times
e^{-2\lambda_{a(1)}\,\chi_{a(1)|d(1)}^{{\rm net}}(y_{2},\vec{b}_{2}|Y,\vec{b})}
\right]
 e^{-2\lambda_{d(1)}\,
 \chi_{d(1)|a(1)}^{{\rm net}}(Y-y_{2},\vec{b}-\vec{b}_{2}|Y,\vec{b})}
-2\lambda_{a(1)}\,
\bar{\chi}_{a(1)|d(1)}^{{\rm gap}}(y_{2},\vec{b}_{2},y_{{\rm gap}}|Y,\vec{b})
\right\} .\label{chi-gap}\end{eqnarray}

It is useful to obtain an alternative representation for $2\bar{\chi}_{a(1)|d(1)}^{{\rm gap}}$,
considering explicitely all couplings of uncut {}``net fans'' to
the {}``handle'' of the diffractively cut {}``net fan'' (see Fig.~\ref{fan-dif-alt-fig}):%
\begin{figure}[t]
\begin{center}\includegraphics[%
  width=12cm,
  height=5cm]{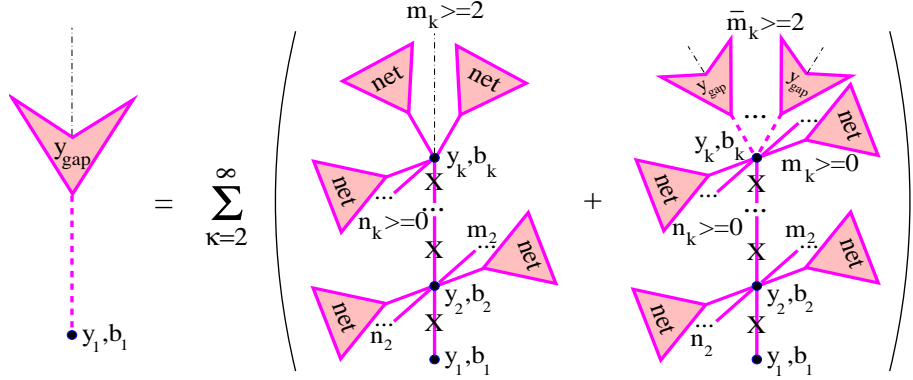}\end{center}
\vspace{-4mm}

\caption{Alternative representation for the contribution $2\bar{\chi}_{a(1)|d(1)}^{{\rm gap}}$
of {}``fan''-like diffractive cuts of {}``net-fan'' diagrams.
The cut pomeron, exchanged between the vertices ($y_{1},b_{1}$) and
($y_{k},b_{k}$), may contain any number $k-2\geq0$ of intermediate
vertices, each one connected to $m_{i}$ projectile and $n_{i}$ target
{}``net fans''; $m_{i},n_{i}\geq0$, $m_{i}+n_{i}\geq1$, $i=2,...,k-1$.\label{fan-dif-alt-fig}}
\end{figure}
\begin{eqnarray}
2\,\bar{\chi}_{a(1)|d(1)}^{{\rm gap}}(y_{1},\vec{b}_{1},y_{{\rm gap}}|Y,\vec{b})=
\sum_{k=2}^{\infty}\frac{G^{k-1}}{\lambda_{a(1)}}\;
\prod_{i=2}^{k}\left[\int_{y_{{\rm gap}}}^{y_{i-1}}\!\! dy_{i}\int\!\! d^{2}b_{i}\;
 G_{\mathbb{PP}}^{\mathbb{P}}(y_{i-1}-y_{i},|\vec{b}_{i-1}-\vec{b}_{i}|)\right]
 \nonumber \\
\times\prod_{j=2}^{k-1}\left[e^{-2\lambda_{a(1)}\,
\chi_{a(1)|d(1)}^{{\rm net}}(y_{j},\vec{b}_{j}|Y,\vec{b})-2\lambda_{d(1)}\,
\chi_{d(1)|a(1)}^{{\rm net}}(Y-y_{j},\vec{b}-\vec{b}_{j}|Y,\vec{b})}-1\right]
e^{-2\lambda_{d(1)}\,
\chi_{d(1)|a(1)}^{{\rm net}}(Y-y_{k},\vec{b}-\vec{b}_{k}|Y,\vec{b})}
\nonumber \\
\times\left[\left(1-e^{-\lambda_{a(1)}\,\chi_{a(1)|d(1)}^{{\rm net}}(y_{k},\vec{b}_{k}|Y,\vec{b})}
\right)^{2}+\left(e^{2\lambda_{a(1)}\,
\bar{\chi}_{a(1)|d(1)}^{{\rm gap}}(y_{k},\vec{b}_{k},y_{{\rm gap}}|Y,\vec{b})}
-1\right)\right. \nonumber \\
\left. \times 
e^{-2\lambda_{a(1)}\,\chi_{a(1)|d(1)}^{{\rm net}}(y_{k},\vec{b}_{k}|Y,\vec{b})}
\right].
\label{chi-gap-alt}\end{eqnarray}

Now we can obtain contributions of diffractive cuts of the diagrams
of Fig.~\ref{enh-full}, using (\ref{chi-gap-alt}) and Fig.~\ref{fan-dif-alt-fig}
to correct for double counting of some graphs in the same manner as
in \cite{ost06} for elastic scattering diagrams. In particular, for
the process of central diffraction, separated from the projectile
and the target by rapidity gaps of sizes larger or equal  to $y_{{\rm gap(1)}}$ and $y_{{\rm gap(2)}}$
correspondingly, we have (see Fig.~\ref{2-gap-fig}):%
\begin{figure}[t]
\begin{center}\includegraphics[%
  width=15cm,
  height=4cm]{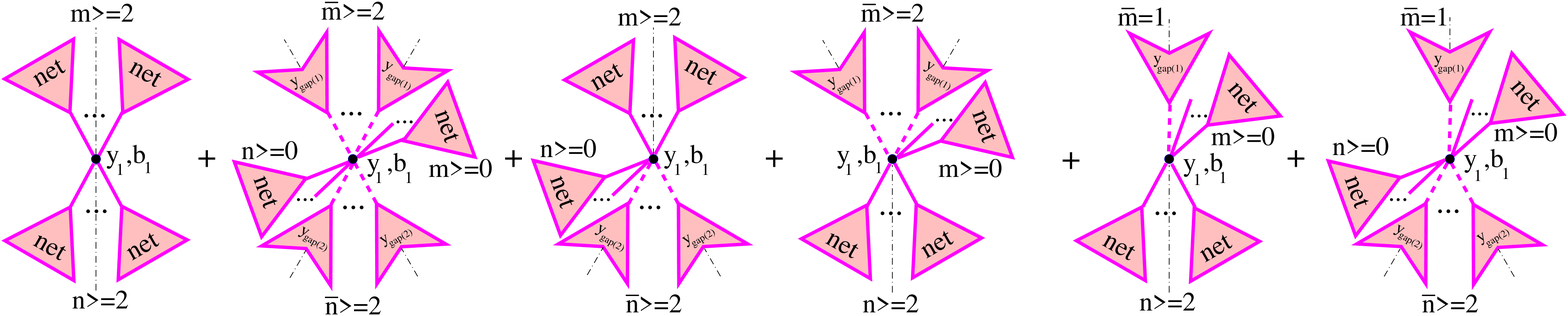}\end{center}
\vspace{-4mm}

\caption{Cut enhanced graphs of double rapidity gap topology.\label{2-gap-fig}}
\end{figure}
\begin{eqnarray}
2\,\chi_{ad(11)}^{{\rm 2-gap}}\!(s,b,y_{{\rm gap(1)}},y_{{\rm gap(2)}})=
\frac{G}{2\lambda_{a(1)}\,\lambda_{d(1)}}\int_{y_{{\rm gap(2)}}}^{Y-y_{{\rm gap(1)}}}\!\! dy_{1}\!
\int\!\! d^{2}b_{1}\nonumber \\
\times\left[\left(1-e^{-\lambda_{a(1)}\,\chi_{a(1)|d(1)}^{{\rm net}}}\right)^{2}+
\left(e^{2\lambda_{a(1)}\,\bar{\chi}_{a(1)|d(1)}^{{\rm gap}}}-1\right)
e^{-2\lambda_{a(1)}\,\chi_{a(1)|d(1)}^{{\rm net}}}\right]\nonumber \\
\times\left[\left(1-e^{-\lambda_{d(1)}\,\chi_{d(1)|a(1)}^{{\rm net}}}\right)^{2}+
\left(e^{2\lambda_{d(1)}\,\bar{\chi}_{d(1)|a(1)}^{{\rm gap}}}-1
-2\lambda_{d(1)}\,\bar{\chi}_{d(1)|a(1)}^{{\rm gap}}\right)
e^{-2\lambda_{d(1)}\,\chi_{d(1)|a(1)}^{{\rm net}}}\right].\label{2-gap}\end{eqnarray}
Here the arguments of the eikonals in the r.h.s.~of (\ref{2-gap})
are understood as 
$\bar{\chi}_{a(1)|d(1)}^{{\rm gap}}=
\bar{\chi}_{a(1)|d(1)}^{{\rm gap}}(Y-y_{1},\vec{b}-\vec{b}_{1},y_{{\rm gap(1)}}|Y,\vec{b})$,
$\bar{\chi}_{d(1)|a(1)}^{{\rm gap}}=
\bar{\chi}_{d(1)|a(1)}^{{\rm gap}}(y_{1},\vec{b}_{1},y_{{\rm gap(2)}}|Y,\vec{b})$,
$\chi_{a(1)|d(1)}^{{\rm net}}=\chi_{a(1)|d(1)}^{{\rm net}}(Y-y_{1},\vec{b}-\vec{b}_{1}|Y,\vec{b})$,
$\chi_{d(1)|a(1)}^{{\rm net}}=\chi_{d(1)|a(1)}^{{\rm net}}(y_{1},\vec{b}_{1}|Y,\vec{b})$.
It is easy to verify that for $y_{{\rm gap(1)}}=y_{{\rm gap(2)}}$
the expression (\ref{2-gap}) is symmetric under the replacement $(a\longleftrightarrow d)$,
which can be made obvious if we expand the projectile diffractively
cut {}``fan'' $\bar{\chi}_{a(1)|d(1)}^{{\rm gap}}$ in the last two graphs
of Fig.~\ref{2-gap-fig} using the relation of Fig.~\ref{fan-dif-alt-fig}.

\begin{figure}[t]
\begin{center}\includegraphics[%
  width=15cm,
  height=4cm]{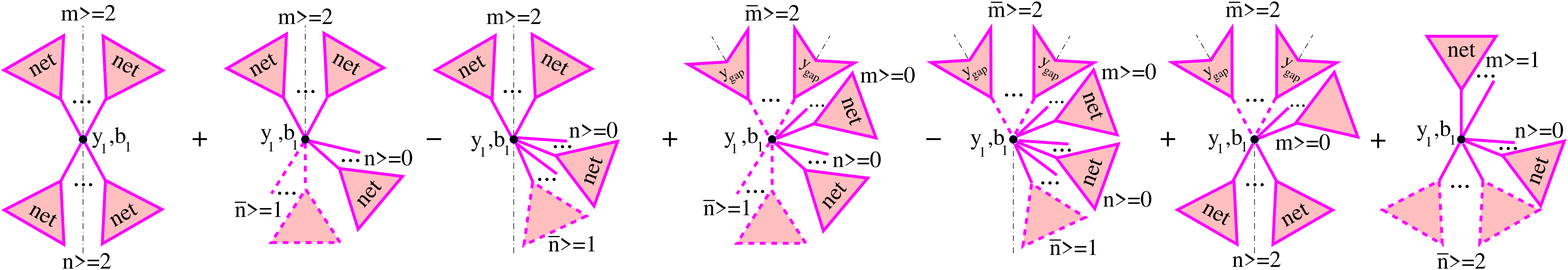}\end{center}
\vspace{-4mm}

\caption{Cut enhanced graphs of target diffraction topology: projectile hadron
is separated from other secondary particles by a large rapidity gap.\label{1-gap-fig}}
\end{figure}

In turn, requiring at least one rapidity gap of size $\geq y_{{\rm gap}}$
between the projectile hadron and the nearest hadron produced after
the gap, we obtain the set of diagrams of Fig.~\ref{1-gap-fig},%
 which gives%
\footnote{Strictly speaking, in the last diagram of 
Fig.~\ref{1-gap-fig}
the size of the rapidity gap is larger than $Y-y_{1}$. For simplicity,
this is neglected here, the effect being negligible for  diffraction
cross sections. In practice, main contributions to single and central
diffraction come from the first three graphs in Fig.~\ref{1-gap-fig}
and from the 1st and the 5th graphs  in Fig.~\ref{2-gap-fig} correspondingly.
Other diagrams are proportional to the third or higher power of the
triple pomeron constant and can be neglected in the region not suppressed
by the elastic from factor (see (\ref{sig-hmd})). This was precisely the
reason to neglect  diffractive cuts of  {}``net fans'', which leaved the
 {}``handle'' of the {}``fan'' uncut.%
}\begin{eqnarray}
2\,\chi_{ad(11)}^{{\rm 1-gap}}\!(s,b,y_{{\rm gap}})=\frac{G}{\lambda_{a(1)}\,\lambda_{d(1)}}
\int_{0}^{Y-y_{{\rm gap}}}\!\! dy_{1}\!\int\!\! d^{2}b_{1}\;\left\{ \left[\left(1-e^{-\lambda_{a(1)}\,
\chi_{a(1)|d(1)}^{{\rm net}}}\right)^{2}\right.\right.\nonumber \\
\left.+\left(e^{2\lambda_{a(1)}\,\bar{\chi}_{a(1)|d(1)}^{{\rm gap}}}-1
-2\lambda_{a(1)}\,\bar{\chi}_{a(1)|d(1)}^{{\rm gap}}\right)
e^{-2\lambda_{a(1)}\,\chi_{a(1)|d(1)}^{{\rm net}}}\right]\left(1-e^{\lambda_{d(1)}\,\tilde{\chi}_{d(1)|a(1)}^{{\rm fan}}-\lambda_{d(1)}\,\chi_{d(1)|a(1)}^{{\rm net}}}\right)\nonumber \\
+\left.\left(1-e^{-\lambda_{a(1)}\,\chi_{a(1)|d(1)}^{{\rm net}}}\right)\left(e^{\lambda_{d(1)}\,\tilde{\chi}_{d(1)|a(1)}^{{\rm fan}}}-1-\lambda_{d(1)}\,\tilde{\chi}_{d(1)|a(1)}^{{\rm fan}}\right)e^{-\lambda_{d(1)}\,\chi_{d(1)|a(1)}^{{\rm net}}}\right\},\label{1-gap}\end{eqnarray}
where the relation (\ref{equiv}) was taken into account and
the arguments of the eikonals in the r.h.s.~of (\ref{1-gap}) are
understood as $\bar{\chi}_{a(1)|d(1)}^{{\rm gap}}=
\bar{\chi}_{a(1)|d(1)}^{{\rm gap}}(Y-y_{1},\vec{b}-\vec{b}_{1},y_{{\rm gap}}|Y,\vec{b})$,
$\chi_{a(1)|d(1)}^{{\rm net}}=\chi_{a(1)|d(1)}^{{\rm net}}(Y-y_{1},\vec{b}-\vec{b}_{1}|Y,\vec{b})$,
$\chi_{d(1)|a(1)}^{{\rm net}}=\chi_{d(1)|a(1)}^{{\rm net}}(y_{1},\vec{b}_{1}|Y,\vec{b})$,
$\tilde{\chi}_{d(1)|a(1)}^{{\rm fan}}=\tilde{\chi}_{d(1)|a(1)}^{{\rm fan}}(y_{1},\vec{b}_{1}|Y,\vec{b})$.

Now, summing over any number but at least one rapidity gap contribution
$2\chi_{ad(11)}^{{\rm 1-gap}}\!(s,b,y_{{\rm gap}})$ and over any number
of elastic re-scatterings, described by the eikonal factor $2\chi_{ad(11)}^{{\rm tot}}(s,b)$
(see (\ref{chi_ad-tot})), selecting in the cut plane elastic intermediate
state for the projectile hadron (cf.~with (\ref{LMD-proj})), and
subtracting central diffraction contribution, we obtain target single
high mass diffraction cross section as\begin{eqnarray}
\sigma_{ad}^{{\rm HMD(targ)}}(s,y_{{\rm gap}})=
\int\!\! d^{2}b\,\left\{ C_{a(1)}^{2}\, C_{d(1)}
\left(e^{2\lambda_{a(1)}\,\lambda_{d(1)}\,
\chi_{ad(11)}^{{\rm 1-gap}}\!(s,b,y_{{\rm gap}})}-1\right)
 e^{-2\lambda_{a(1)}\,\lambda_{d(1)}\,\chi_{ad(11)}^{{\rm tot}}(s,b)}
 \right.\nonumber \\
\left.-C_{a(1)}^{2}\, C_{d(1)}^{2}
\left(e^{2\lambda_{a(1)}\,\lambda_{d(1)}\,
\chi_{ad(11)}^{{\rm 2-gap}}\!(s,b,y_{{\rm gap}},0)}-1\right)
e^{-2\lambda_{a(1)}\,\lambda_{d(1)}\,\chi_{ad(11)}^{{\rm tot}}(s,b)}
\right\}.\label{sig-hmd}\end{eqnarray}
Here the central diffraction term in the 2nd line of (\ref{sig-hmd})
is obtained summing over any number but at least one double gap contribution
$2\chi_{ad(11)}^{{\rm 2-gap}}\!(s,b,y_{{\rm gap}},0)$ (for any size
of the second gap) and over any number of elastic re-scatterings and
selecting in the cut plane elastic intermediate states for both hadrons.
Projectile single high mass diffraction cross section $\sigma_{ad}^{{\rm HMD(proj)}}(s,y_{{\rm gap}})$
is obtained via the replacement $(a\longleftrightarrow d)$ in the
r.h.s.~of (\ref{sig-hmd}).

\end{document}